\documentclass[9pt]{style}

\usepackage[utf8]{inputenc}
\usepackage{import}
\usepackage{array}
\usepackage{tabularx}
\usepackage{wrapfig}
\usepackage{textcomp}
\usepackage{amsmath}
\usepackage{graphicx}
\usepackage{esint}
\usepackage{subscript}
\usepackage{rotating}
\usepackage{float}
\usepackage{graphicx,kantlipsum,setspace}
\usepackage{tikz}
\usepackage{tikzscale}
\usepackage{pgfplots}
\usepackage{lmodern}
\usepackage{lscape}

\usepackage{caption}
\usepackage{multirow}
\usetikzlibrary{arrows}
\makeatletter
\def\ps@pprintTitle{%
 \let\@oddhead\@empty
 \let\@evenhead\@empty
 \def\@oddfoot{\centerline{\thepage}}%
 \let\@evenfoot\@oddfoot}

\@ifundefined{definecolor}{\usepackage{color}}{}
\definecolor{Wine}{rgb}{0.612, 0.192, 0.388}

\usetikzlibrary{matrix,chains,positioning,decorations.pathreplacing,arrows,arrows.meta}
\usetikzlibrary{matrix,positioning,calc}
\usetikzlibrary{decorations.pathmorphing}

\xdefinecolor{c1}{HTML}{4ED99C}
\xdefinecolor{Ecol}{HTML}{FF6868}
\xdefinecolor{Icol}{HTML}{3366cc}

\usepackage{marginnote}
\usepackage{lineno}

\@ifundefined{showcaptionsetup}{}{%
 \PassOptionsToPackage{caption=false}{subfig}}
\usepackage{subfig}
\makeatother

\newcommand\inputpgf[2]{{
\let\pgfimageWithoutPath\pgfimage
\renewcommand{\pgfimage}[2][]{\pgfimageWithoutPath[##1]{#1/##2}}
\input{#1/#2}
}}

\newlength\figureheight
\newlength\figurewidth

\title{Gap Junction Plasticity can Lead to Spindle Oscillations.}

\author[1,2]{Guillaume Pernelle}
\author[1]{Wilten Nicola}
\author[1]{Claudia Clopath}
\affil[1]{Department of Bioengineering, Imperial College London}
\affil[2]{Corresponding author: g.pernelle14@imperial.ac.uk}{}

\pgfplotsset{compat=1.14}
\begin{document}

\maketitle

\begin{abstract}

Patterns of waxing and waning oscillations, called spindles, are observed in multiple brain regions during sleep. Spindle are thought to be involved in memory consolidation. The origin of spindle oscillations is ongoing work but experimental results point towards the thalamic reticular nucleus (TRN) as a likely candidate. The TRN is rich in electrical synapses, also called gap junctions, which promote synchrony in neural activity. Moreover, gap junctions undergo activity-dependent long-term plasticity. We hypothesized that gap junction plasticity can modulate spindle oscillations. We developed a computational model of gap junction plasticity in recurrent networks of TRN and thalamocortical neurons (TC). We showed that gap junction coupling can modulate the TRN-TC network synchrony and that gap junction plasticity is a plausible mechanism for the generation of sleep-spindles. Finally, our results are robust to the simulation of pharmacological manipulation of spindles, such as the administration of propofol, an anesthetics known to generate spindles in humans.

\end{abstract}

\section{Author summary.}
During non-rapid-eye-movement sleep, neurons tend to become active together for about one second, before desynchronising. This waxing and waning pattern of neuronal activity is called spindle oscillations and is thought to be important for memory formation. Spindle oscillations travel throughout the cortex and are thought to originate from a small region within the thalamus, the thalamic reticular nucleus (TRN). The TRN has singular characteristics: it contains only inhibitory neurons and they are coupled with electrical synapses. Many studies highlight the role of electrical synapses between inhibitory neurons to generate oscillations. Moreover, they have been shown to be plastic, their strength can be altered by the neuronal activity. Therefore, the plasticity in electrical synapses could act as a mechanism to generate spindle oscillations. We developed a computational model of plasticity of electrical synapses in a network of thalamic neurons. We first show that electrical synapses can be a lead factor of oscillations within the thalamus. Then, we show that spindle oscillations can be generated by the gap junction plasticity.

\section{Introduction.}

Spindle oscillations are waxing and waning patterns which are characterized by field potentials oscillating at 7-15 Hz and occurring every 5-15 seconds (\cite{Steriade1993,McCormick1997a}).
They can be observed during sleep and their absence can be a marker of schizophrenia (\cite{Ferrarelli2010}).

The thalamic reticular nucleus (TRN) is thought to be involved in the genesis of sleep spindles (\cite{VonKrosigk1993,Lee2013,Halassa2014}). The TRN consists of inhibitory neurons and surrounds the dorsal thalamus like a shell (\cite{Houser1980,Pinault2004}). Interestingly, spindle oscillations are observed in thalamus after decortication (\cite{Morison1945,Timofeev1996}) or cortical deafferation \cite{Steriade1987,Lemieux2014} and they are not observed in the isolated cortex (\cite{Fuentealba2004}). Moreover, they are also observed in isolated TRN \textit{in vitro} (\cite{Deschenes1985}).

The "classical" model postulates that sleep spindles arise from the interactions between feedfoward inhibition from the TRN onto thalamo-cortical (TC) neurons and feeback excitation from the TC to the TRN (\cite{Bal1995,Cox1996,Timofeev2001,Timofeev2013}).
Computational models also suggest that the mutual inhibition from TRN and TC neurons could generate spindle activity in the TRN (\cite{Wang1993,Destexhe1994}). Another possible factor in the existence of thalamic sleep spindles is the two distinctive firing patterns that TRN and TC neurons exhibit: sequences of tonic action potentials and high frequency bursts of action potentials (\cite{Jahnsen1984,Mulle1986,Avanzini1989,Contreras1992,Contreras1993,Bazhenov2000a,Sherman2001,Sun2012}). The tonic firing mode can be observed during wakeful states (\cite{McCormick1990}) while burst firing mode can be observed during spindles (\cite{Halassa2011}). Modelling studies suggest that the transition from tonic to bursting firing mode can be associated with the transition from asynchronous to synchronous network regime (\cite{Postnova2007,Shaffer2017}).

The TRN is particularly dense in interneurons coupled with electrical synapses (gap junctions) \cite{Landisman2002}. Interestingly, gap junctions have been shown to promote activity synchronization (\cite{Traub2001,Pfeuty2003,Kopell2004,Connors2004a,Tchumatchenko2014}) and some models consider their involvement in spindle oscillations (\cite{Landisman2002,Lewis2003,Long2004}). Experimental results show that administration of halothane, a gap junction blocker, in TRN of decorticated cats lead to a reduction of spindle activity (\cite{Fuentealba2004}).
Moreoever, it has been show that gap junctions are plastic (\cite{Cachope2007a,Wang2015a,Turecek2014,Turecek2016,Coulon2017}). High frequency stimulation of a pair of TRN inhibitory neurons coupled by gap junctions leads to long-term depression of the gap junctions (gLTD) (\cite{Haas2011}).

Given the probable involvement of gap junctions in the synchronous pattern of the TRN, and the existence of activity-dependent gap junction plasticity, we hypothesized that gap junction plasticity could act as a mechanism to generate spindle oscillations. 
To test this, we developed a model of TRN and TC neurons. Consistent with our previous work in cortex (\cite{Tchumatchenko2014,Pernelle2017}), the gap junction coupling strength of TRN neurons modulates the TRN-TC network synchrony (\cite{Tchumatchenko2014,Pernelle2017}). Then, we show that gap junction plasticity can lead to the emergence of waxing and waning oscillations that have the same characteristics as sleep spindles observed during stage 2 of non-rapid-eye-movement sleep (NREM). Finally, we show that our model allows for manipulation of spindles in agreement with experimental results.

\section{Results}
\subsection{Gap junction coupling between TRN inhibitory neurons promotes synchrony in the TRN-TC network.}
In order to understand the effect of gap junction plasticity on spindle activity, we consider first a static network of inhibitory TRN neurons, fully connected via chemical and electrical synapses (Figure 1A). 
The Izhikevich model of rat TRN neurons was used to reproduce the firing and bursting modes (\cite{Izhikevich2007}). 
As previously reported \cite{Tchumatchenko2014,Pernelle2017}, we observed a synchronous regular (SR) regime for strong gap junction coupling and external network drive, and an asynchronous irregular (AI) regime for low levels of both (Figure 1B). 
With stronger gap junction coupling, the inhibitory neurons merge their electrical properties and synchronize together. This is due to the current contribution of the gap junction coupling which is proportional to the difference of voltages between the coupled neurons. 
Moreover, spiking inhibitory neurons emit an excitatory spikelet in the coupled inhibitory neurons, on top of the inhibition provided by the inhibitory chemical synapses.
We observe that the TRN neurons exhibit the two firing modes characterizing TRN neurons. 
Tonic firing mode is mostly observed in the AI regime, while burst firing mode is mostly observed in the SR regime (Figures 1C and S1). 
To complete our thalamo-cortical model, we connect in an all-to-all fashion the TRN neurons to TC neurons that are excitatory (Figure 1D). 
We observe that the excitatory neurons follow the activity patterns generated by the TRN neurons (Figures 1E-I and S2).

To summarize, the synchrony of the TRN-TC network increases with the gap junction coupling strength and input drive. 
To understand the effect of gap junction plasticity on spindle activity, we model the plasticity, similarly as in \cite{Pernelle2017}.

\subsection{Gap junction plasticity dynamics drive spindle activity.}
\cite{Haas2011} reported the first experimental evidence of gap junction activity-dependent long-term depression. After identifying pairs of neurons coupled by gap junctions, they repetitively provoked bursts during five minutes in one or both neurons and they observed long-term depression (gLTD) of the coupling coefficient.
We therefore used a model of gap junction plasticity that is activity-dependent (developed in \cite{Pernelle2017}). 
The frequency of the spiking activity determines the direction of the plasticity. 
We hypothesize that sparse firing lead to gap junction long-term potentiation (gLTP), while repetitive burst firing leads to gLTD (Figure 2A). 
This rule allows for the emergence of spindle activity. 
From the AI regime, sparse firing leads to an increase in the gap junction coupling strength, which drives the TRN network in the SR regime (waxing phase, Figure 2B). Then, due to the sustained bursting of TRN neurons, depression of the gap junction follows and the network crosses back the bifurcation between AI and SR regime and enters the AI regime again (waning phase, Figure 2B). This network dynamic creates a waxing-and-waning mechanism that characterize sleep spindles.

\subsection{Gap junction coupling resembles a chaotic attractor.}
The network dynamic, previously observed for a single spindle event, repeats itself irregularly (5-10 seconds) and has an irregular duration (0.5-2 seconds) (Figures 3A and S3B-D). 
The spindle activity, driven by the gap junction coupling strength, is not purely periodic. 
We observe indeed a variation in the spindle duration, power and intervals between spindles. The delay-dimension embedded phase portrait bares a strong resemblance to a chaotic attractor (Figure 3B). Moreover, from identical initial conditions, small perturbations to the gap junction coupling will cause the dynamical system to diverge at an exponential rate before saturation, which is distinctive of a chaotic system (Figure 3C-D).

\subsection{TRN activation suppresses spindle activity.}
To test our model, we looked at the experimental literature for perturbation experiments, and then tested whether our model is consistent with the data. In particular, \cite{Lewis2015} observed a signification reduction of spindle activity during stimulation of the TRN. We wondered if our model would exhibit a similar effect. We started from conditions that allow spindle activity as previously described (Figure 4A). 
With a TRN external drive of 40 pA, repetitive patterns of waxing-and-waning oscillations are observed. 
When increasing the drive to the TRN to 50 pA, the TRN-TC network becomes asynchronous and the spindle activity is suppressed (Figure 4B). 
In order to understand this effect, we investigated the spiking activity and the oscillation power, which influence the gap junction plasticity. 
When superimposing lines of iso-power (Figure 4C) and iso-activity (Figure 4D), we observe that, while at 40 pA they coincide, they start to diverge when activating the TRN. 
When the drive $\nu$ is 50 pA, the amount of bursting activity, and therefore gLTD, is higher for the same level of network oscillation. 
This implies that it becomes more difficult for the gap junction coupling to reach values that are associated with the SR regime. Thus, the gap junction plasticity stabilizes at a fixed point outside the SR regime, which results in the absence of spindle oscillations. 
We observed that spindles can be suppressed by increasing the amount of bursting and creating a plasticity fixed point in the AI regime. Similar conditions may be observed in awake animals, where no spindle activity is observed.

\subsection{Manipulation of spindle activity with propofol.}
As a second test, we wondered whether our model is consistent with the effects of propofol on spindles. Propofol is an anesthetics that promotes spindle activity in humans (\cite{Murphy2011}). It targets $GABA_A$ receptors and prolongs the duration of GABAergic inhibitory postsynaptic currents (IPSCs) (\cite{MacIver1991,Orser1994,Bai1999}). 
To simulate this effect, we increase the time constant of chemical inhibitory synapses $\tau_I$. Starting from a network in the asynchronous regime (Figure 4B), once $\tau_I$ crosses a threshold which depends on the external TRN drive (Figure S3A), the TRN exhibits spindle activity (Figure 5A). 
On the updated activity diagram ($\tau_I$ = 20 ms), when superimposing the line of iso-power (obtained from a power diagram for $\tau_I$ = 20 ms) to the line of iso-activity, we observed that the iso-activity line stays to the right of the iso-power line unlike for a smaller $\tau_I$ (Figure 4D). 
Increasing the duration of GABAergic IPSCs leads to a decrease in bursting activity for the same level of network oscillation. 
Therefore, the gap junction coupling can reach values associated with the SR regime. The gap junction coupling oscillatory cycle through the bifurcation between the AI and SR regime is restored and spindle activity is observed.
This effect increases with $\tau_I$ and the power, duration, and inter-spindle-interval increase with $\tau_I$ (Figures 4C-E).

To summarize, mimicking the effect of propofol on the dynamic of inhibitory synapses can generate spindle activity through gap junction plasticity. 


\newpage
\section{Discussion.}
To summarize, our modeling study tested whether gap junction plasticity can generate spindle oscillations in the thalamus. 
Our findings suggest that gap junctions between TRN neurons can regulate the synchrony in both the TRN and TC. 
Moreover, we showed that, depending on the dynamics of inhibitory synapses, gap junction plasticity can generate spindle oscillations. 
Gap junctions generate patterns of waxing-and-waning spindle oscillations in the population activity, with their strength oscillating in a non purely periodic fashion around the bifurcation between asynchronous irregular (AI) and synchronous regular (SR) regimes. In the AI regime, sparse firing leads to potentiation of the gap junctions, which drives the network towards the SR regime. 
However, once in the SR regime, oscillations emerge, which lead to repetitive bursting of TRN neurons. This trigger gap junction depression and drives the network back to the AI regime. 
This alternation between AI and SR regimes can generate spindle like activity. This mechanism is robust to the dynamics of inhibitory synapses and the external drive to the network. 
For fast synapses or strong TRN activation, the gap junction plasticity reaches a steady-state in the AI regime and no spindles are observed. 
Slowing down the dynamics of inhibitory synapses or reducing the TRN activation promotes the emergence of spindle oscillations. 
Therefore, gap junction plasticity is a potential mechanism in generating spindle oscillations.

While there is no experimental data yet on activity-dependent LTP of gap junctions, recent studies suggest that gap junction gLTP and gLTD share similar calcium-regulated pathways (\cite{Wang2015b,Szoboszlay2016,Sevetson2017}). Therefore we designed a rule for activity-dependent gLTP consistent with Haas' results (\cite{Haas2011}). We assumed that sparse firing would lead to gLTP. Our results do not depend stricly on this assumption. Indeed, in a previous study we reported that passive gLTP mechanism would yield similar results (\cite{Pernelle2017}).

Our computational model shows that gap junction plasticity could be involved in the genesis of spindle oscillations. 
The spindle activity generated by our model bares the hallmark of chaotic activity. This corroborates modeling work of gap junction coupling, but between excitatory TC neurons (\cite{Hughes2004,Ermentrout2006}). However, further numerics are required to characterize this system, which could be an interesting future work. Unfortunately, the dynamics of the plasticity mechanism makes these systems computationally intensive to simulate, requiring very long simulation times.  
Furthermore, \cite{Fuentealba2004} observed a reduction of the spindle activity \textit{in vivo} when blocking gap junction with administration of halothane in TRN of decorticated cats. Those results are consistent with our model, where gap junctions are the leading factor in the the emergence of oscillations. 

Alternative models for spindle generation have been proposed. For example, a classical model takes the minimal substrate for spindles to be the interaction between TRN and TC neurons. Bursts from TRN inhibitory neurons trigger bursts of excitatory thalamic relay cells, which then excite the TRN neurons and the cycle continues (\cite{Steriade1984,Steriade1985,VonKrosigk1993,Bal1995,Cox1996,Timofeev2001,Bonjean2011,Timofeev2013}). 
With our model however, spindles can be generated from the gap junction plasticity in isolated TRN.
This is consistent with observations \textit{in vitro} from \cite{Deschenes1985} or the \textit{in vivo} study from \cite{Steriade1987}.
Those mechanisms for spindle generation are not mutually exclusive, they could act in concordance and guarantee the robustness of this mechanism or they could promote different types of spindles, such as the slow and fast spindles (\cite{Andrillon2011}).

Spindle oscillations appear involved in memory formation and consolidation (\cite{Bazhenov1998,Gais2000,Rosanova2005,Werk2005,Astori2013,Latchoumane2017,Xia2017}). Thus, better understanding the potential implications of gap junction plasticity in spindle generation could further our understanding on memory formation. 

To conclude, we showed that gap junction plasticity can modulate the synchrony in the thalamus and generate spindle oscillations. Finally, our model is consistent with experimental perturbation studies.


\newpage
\section{Methods.}
We consider a network with $N_I$ TRN inhibitory neurons and $N_E$ TC excitatory neurons with all-to-all connectivity (Figure 1).
Inhibitory neurons are modelled by an Izhikevich model and excitatory neurons by a leaky integrated-and-fire model (LIF) (\cite{Izhikevich2003,Izhikevich2007}). 
Equations are integrated using Euler's method with a time-step of 1 millisecond. This large time-step was used as numerical simulations of gap junction plasticity are very computationally intensive. However, we performed tests with smaller time-steps (up to 0.02 ms) and we observed similar neuron and network dynamics.
Inhibitory neurons are connected by both electrical and chemical synapses, whereas excitatory neurons have only chemical synapses. 
The model is written in Python and is available on github: $https://github.com/gpernelle/spindles\_with\_gap\_junctions$.

\subsection*{Neuron models.} \label{s:izh}
We model TRN neurons with Izhikevich type neuron models for rat TRN neurons (\cite{Izhikevich2004,Izhikevich2007}). 
This model reproduces the tonic spiking and bursting that is typical of TRN neurons.  
The voltage $v$ follows 
\begin{equation} 
	\tau_v \dot{v} =  k_v (v-v_{ra})(v-v_{rb}) - k_u u + R I,
\end{equation}
\begin{equation} 
	\tau_u \dot{u} = a [ c (v - v_{rc}) - u], 
\end{equation}
combined with the following conditions,
	\begin{equation}
		\mbox{
			if $v \geq v_{peakTRN}$,
			then}
			\left\lbrace{ \begin{array}{l} v \leftarrow v_{resetTRN} \\ u \leftarrow  u+b,
\end{array}} \right.  
\end{equation}

\begin{equation}
		\mbox{and if $v \leq v_{switch} $, then c = 10 else c = 2.}
\end{equation}
where $\tau_v$ is the membrane time constant, $v_{ra}$ is the membrane resting potential, $v_{rb}$ is the membrane threshold potential, $k_v$ is a voltage parameter, $k_u$ is the coupling parameter to the adaptation variable $u$, $R$ is the resistance and $I$ is the current. 
The adaptation variable $u$ represents a membrane recovery variable, accounting for the activation of K$^+$ ionic currents and inactivation of Na$^+$ ionic currents. It increases by a discrete amount $b$ every time the neuron is spiking and its membrane potential crosses the threshold $v_{peakTRN}$.
It provides a negative feedback to the voltage $v$.
$\tau_u$ is the recovery time constant, $a$ and $c$ are coupling parameters, $v_{resetTRN}$, $b$ and $v_{rc}$ are voltage constants. We slightly modified Izhikevich's model. In our model, $v_{resetTRN}$ was reduced from -55 mV to -60 mV to decrease the spontaneous bursting activity that could be observed in the asynchronous irregular regime. The parameter $v_{switch}$ was reduced from -65 mV to -70 mV to slightly decrease the spindle oscillation frequency.

To model TC excitatory neurons, we chose a leaky integrate-and-fire model, 
\begin{equation} \tau_m  \dot{v} =  -v + R_m I,
\end{equation}
where $\tau_m$ is the membrane time constant, $v$ the membrane potential, $I$ the current and $R_m$ the resistance.
Spikes are characterized by a firing time $t_f$ which corresponds to the time when $v$ reaches the threshold $v_{threshTC}$.
Immediately after a spike, the potential is reset to the reset potential $v_{resetTC}$.

\subsection*{Network.}\label{sec:network}

In the single network model (Figures 1 and 2), each neuron is connected to all others by chemical synapses. Additionally, inhibitory neurons are connected via  electrical synapses to all other inhibitory neurons, as in~\cite{Tchumatchenko2014}. 
Thus, the current each individual neuron $i$ receives can be decomposed in four components 
\begin{equation} 
I _{i}(t)= I^{spike}_{i}(t) +I^{gap}_{i}(t) + I^{noise}_{i}(t) + I^{ext}_{i}(t), 
\end{equation}
where $I^{spike}_{i} = I^{chem}_{i} + I^{elec}_{i}$ is the current coming from the transmission of a spike via electrical (i.e. spikelet) and chemical synapses, $I^{gap}_{i}$ is the sub-threshold current from electrical synapses (for inhibitory neurons only), $I^{noise}_{i}$ is the noisy background current and $I^{ext}_{i}$ characterizes the external current.  
The current due to spiking $I^{spike}_{i}$ on excitatory neurons is given by 
\begin{equation}
	I^{spike}_{i}(t) = W^{IE} 
	\sum_{	
	\substack{
				j=1\\
				j \neq i}
		}^{N_I}
		\sum_{
			t_{jk}<t
		} 
			\exp{\left(
				-\frac{t-t_{jk}}{\tau_{I}}
				\right)}
			\\
			+
	W^{EE}  \sum_{	
	\substack{
				j=1\\
				j \neq i}
		}^{N_E}
		\sum_{
			t_{jk}<t
			} 
				\exp{\left(-\frac{t-t_{jk}}{\tau_{E}}\right)}.
				\end{equation}
				
The current $I^{spike}_{i}$ into inhibitory neurons are
\begin{equation}
					I^{spike}_{i}(t) =
		\sum_{	
	\substack{
				j=1\\
				j \neq i}
		}^{N_I}
		\sum_{
			t_{jk}<t
		} W^{II}_{ij} 
			\exp{\left(
				-\frac{t-t_{jk}}{\tau_{I}}
				\right)}
			\\
			+
	W^{EI} 	\sum_{	
	\substack{
				j=1\\
				j \neq i}
		}^{N_E}
		\sum_{ 
			t_{jk}<t
			} 
				\exp{\left(-\frac{t-t_{jk}}{\tau_{E}}\right)},
\end{equation}
where $W^{\alpha\beta}$ is the coupling strength from population $\alpha$ to population $\beta$ with $\{\alpha,\beta\} = \{E,I\}$.  
Finally, $W^{II}_{ij} = W^{II,c} + W^{II,e}_{ij}$ is the inhibitory to inhibitory coupling between neuron $i$ and $j$, consisting of the chemical synaptic strength $W^{II,c}$ and $W^{II,e}_{ij}$ the electrical coupling for supra-threshold current, also called the spikelet. 
We model the contribution of the spikelet as a linear function of the gap junction coupling $W^{II,e}_{ij} = k_{spikelet} * \gamma_{ij}$, where $\gamma_{ij}$ is the gap junction coupling between neurons $i$ and $j$.
Note that $W^{EE}$, $W^{EI}$, $W^{IE}$, $W^{II,c}$ are identical among neurons, but $W^{II}_{ij}$ varies as the spikelet contribution depends on the coupling strengths $\gamma_{ij}$, which can be plastic.

We represent the post-synaptic potential response to a chemical or electrical spike with an exponential of the form 
$\exp{\left(-\frac{t-t_{jk}}{\tau_{\alpha}}\right)}$ for $t>t_{jk}$. The excitatory and inhibitory synaptic time constants are $\tau_{E}$ and $\tau_{I}$ respectively and $t_{jk}$ represents the $k^{th}$ firing time of neuron $j$.  	

In between spikes, for every pair of inhibitory neurons $i,j$, the gap junction
mediated sub-threshold current $I^{gap}_{i}$ is characterized by

\begin{equation}
		I^{gap}_{i}(t) =
		\sum_{	
	\substack{
				j=1\\
				j \neq i}
		}^{N_I}
		I^{gap}_{ij}(t) =
		\sum_{	
	\substack{
				j=1\\
				j \neq i}
		}^{N_I}
		\gamma_{ij} (V_{j}(t) - V_{i}(t)),
\end{equation} 
where $\gamma_{ij}$ is the gap junction coupling between inhibitory neurons $i$ and $j$ of respective membrane
potential $V_{i}$ and $V_{j}$. In our model, we suppose that gap junctions are symmetric with $\gamma_{ij} = \gamma_{ji}$. 
Gap junctions are initialized following a log-normal distribution with the location parameter $\mu_{gap}=1+ln(\gamma / N_I)$ and the scale parameter $\sigma_{gap}=1$.

Neurons also receive the current $I_{noise}$ which is a colored Gaussian noise
with mean $\nu_{I}$, standard deviation $\sigma_{I}$ and $\tau_{noise}$ the time
constant of the low-pass filtering

\begin{equation} 
	\tau_{noise}\dot{s}(t)=-s(t)+\xi(t) 
\end{equation} 
and
\begin{equation} 
	I^{noise}(t) = \sqrt{2\tau_{noise}}s(t)\sigma_{I} + \nu_{I},
\end{equation} 
with $\xi$ is drawn from a Gaussian distribution with unit standard deviation and zero mean.

\subsection*{Plasticity model of gap junctions.} 
Our plasticity model is decomposed into a depression ${\gamma}^{-}$ and a potentiation term ${\gamma}^{+}$, developed in \cite{Pernelle2017}.

\subsection*{gLTD: depression of the electrical synapses for high frequency activity.}  \label{sec:LTD}

To model the gap junction long-term depression (gLTD) measured by \cite{Haas2011} following high frequency activity in one or both neurons, we create a second order filter, composed of two low-pass filters. We first defined a variable $b_{i}$ which is a low-pass filter of the spikes of neuron $i$

\begin{equation}\label{eq:bursting} 
	\tau_{b} \dot{b_{i}}(t) = -b_{i}(t) + \tau_{b} \sum_{t_{ik}<t}  \delta(t-t_{ik}), 
\end{equation}
where $\delta$ is the Dirac function and $\tau_{b}=8$ ms is the time constant.  
Then we defined a variable $q_i$ which is a low-pass filter of $b_i$, that characterizes if the bursting is sustained, 

\begin{equation}
	\tau_{q} \dot{q_{i}}(t) = -q_{i}(t) + p_{i}(t) , 
\end{equation}
where $\tau_q$ is the filter time constant. Then depression is applied when $q_i$ is over a threshold $\theta_q = 0.3$.

In our model, we consider that the individual electrical coupling coefficient $\gamma$ between neurons are non-directional. Therefore, when the interneurons show sustained bursting activity, the gap junctions undergo depression,

\begin{equation}
\label{eq:ltd} \dot{\gamma}^{-}_{ij}(t)  = \dot{\gamma}^{-}_{ji}(t)  = -\alpha_{LTD}  [H(q_i(t)-\theta_{q}) + H(q_j(t)-\theta_{q})] , 
\end{equation} 
where $\alpha_{LTD}$ is the learning rate and $H$ is the Heaviside function that returns 1 for positive arguments and 0 otherwise.

We fit $\alpha_{LTD}$ to the data by implementing the stimulation protocol used in~\cite{Haas2011}, as reported in \cite{Pernelle2017}.

\subsection*{gLTP: potentiation of the electrical synapses for low frequency activity.} \label{sec:LTP} 
If gap junctions were only depressed, they would decay to zero after some time. Therefore, there is a need for gap junction potentiation. However, no activity dependent mechanism has been reported yet in the experimental literature, but \cite{Cachope2012,Wang2015a,Sevetson2017,Debanne2017} suggest that the calcium-regulated mechanisms leading to long-term depression could be involved in potentiation as well. 

To model potentiation, we use the same soft-bound rule as presented in \cite{Pernelle2017},
\begin{equation} \label{eq:ltp-soft} 
	\dot{\gamma}^{+}_{ij}(t) =\dot{\gamma}^{+}_{ji}(t)  =  \alpha_{LTP}  \left(\frac{\gamma_b - \gamma_{ij}(t)}{\gamma_b}\right) [\mathrm{sp}_i(t) + \mathrm{sp}_j(t)]. 
\end{equation}
where $\alpha_{LTP}$ is the learning rate and $\mathrm{sp}_i(t) = \sum_{t_{ik} < t} \delta(t-t_{ik})$.

\subsection*{Quantification of network spiking activity.}

To estimate the plasticity direction for different values of external input $\nu$ and gap junction strength $\gamma$, we observe the activity of the network (without plasticity) in a steady state over a duration $T = 6$ s.
For a chosen tuple $(\nu;\gamma)$, we average over time and over neurons the bursting and spiking activity
\begin{equation}
\label{eq:avg1} 
	A_{bursting} = \frac{1}{T}\int_0^T \frac{1}{N_{I}} \sum_{i=1}^{N_I} [H(b_i(t)-\theta_{burst}) ]dt
\end{equation} and 

\begin{equation}
\label{eq:avg2} 
	A_{spiking} = \frac{1}{T}\int_0^T  \frac{1}{N_{I}} \sum_{i=1}^{N_I} \mathrm{sp}_i(t) dt .
\end{equation} 
Then, we explore the values of the ratio of bursting over spiking activity 
\begin{equation}
	\mathrm{ratio} = \frac{A_{bursting}}{A_{spiking}}
\end{equation}
as function of the coupling coefficient $\gamma$ and of the mean external input $\nu$ over the
parameter space $\mathcal{P}_1 = [0;\gamma_{max}] \times [0;\nu_{max}]$.

\subsection*{Quantification of oscillation power and frequency.}\label{sec:fourier}

To quantify the frequency and the power of the oscillations in the neuronal activity, we perform a Fourier analysis
of the population activity $r$ which  we define as the sum of neuron spikes within a population, during the time step $dt$ 

\begin{equation}\label{eq:PA}
 r(t) = \frac{1}{dt} \frac{1}{N_I}\int_{t}^{t+dt} \sum_{i=1}^{N_I} \sum_{t_{ik} < t}  \delta(u-t_{ik}) du.
 \end{equation}

We compute a Discrete Time Fourier
Transform (DFT) and extract the power and the frequency of the most represented frequency in the Fourier
domain.
The formula defining the DFT is
\begin{equation}
	\hat{r}_k =  \sum_{n=0}^{N-1} r_n \exp{\left(-{i 2\pi k \frac{n}{N}} \right)}
	\qquad
	k = 0,\dots,N-1.
\end{equation}
where the ${r_n}$ sequence represents $N$ uniformly spaced time-samples of the population activities.
We measure the amplitude of the Fourier components $\hat{r}_k$ for $k= 1..N/2$ (because the
Fourier signal is symmetric from $N/2$ to $N$). We identify the maximal one,
its associated frequency $f_{max} = \frac{k}{N}$ and its power $P = (|\hat{r}_k|/N)^2$.

\subsection*{Spectrogram of the population activity.}
To compute the spectrogram of the population activity, the data are split in $NFFT$ length segments and the spectrum of each section is computed. The Hanning windowing function is applied to each segment, which overlap of the 1900 ms, and the mean is removed.

\subsection*{Quantification of spindle power, duration and inter-spindle intervals.}
To quantify the spindle properties, we filter the spectrogram of population activity in the [5-15] Hz frequency band. The beginning $B_i$ and end $E_i$ of each spindle is determined by the rising slope and falling slope of the power crossing the threshold $\theta_{spindle}$. The spindle duration is the difference between falling and rising times $d_i = E_i - B_i$ and the intervals between spindles is the difference between consecutive rising time $\Delta_i = E_{i+1} - E_i$.

\subsection*{Estimation of the largest Lyapunov's exponent.}

The largest Lyapunov's exponent is a measure of the rate of divergence of a system after a perturbation. 
Once the spindle oscillations begin, we perturb the gap junctions. Between different realizations, only the perturbation differs, by setting a different random seed while drawing the perturbation matrix. The perturbation is drawn from a uniform distribution $\mathcal{U}(0;0.1 * \gamma_m)$, where $\gamma_m$ is the mean gap junction coupling strength at the time of the perturbation ($t$ = 10 s).

We then proceed to estimate the rate of divergence between realizations from the perturbation time
\begin{equation}
	e(t) = \frac{1}{N(N-1)} \sum_{i \neq j} |\gamma_{m,i}(t) - \gamma_{m,j}(t)| 
\end{equation}
where $\gamma_{m,i}$ and $\gamma_{m,j}$ are the mean gap junction coupling of realizations $i$ and $j$.

We then perform a linear regression on the logarithm of the mean distance $log(e)$, from the perturbation time until the distance saturates, about 30 seconds after the perturbations. The slope $\lambda$ gives us an estimate of the largest Lyapunov's exponent.

\newpage
\subsection*{Parameters.}
We list in Table 1 the parameters used for our simulations.

\begin{table}[htbp]  
\caption {} \label{tab:1}

\begin{tabular}[t]{l l}

	\begin{tabular}[t]{l}
	\textbf{TRN Interneurons}  \\
	\begin{tabularx}{.4\linewidth}{l r}
	$\tau_{I}$	&	10 ms	\\
	$\tau_I$ for Fig. 3 and 4 	& [10-30] ms \\
	$\tau_v$	&	40 ms \\
	
    $\tau_u$	&	1/0.015 ms \\
	$R$ &	0.6 $\Omega$	\\
	$k_u$ &	1 $\Omega$	\\
	$k_v$ &	0.25 $\Omega$	\\
	$v_{ra}$ & -45 mV	\\
	$v_{rb}$ &	-65 mV	\\
	$v_{rc}$	& -65 mV	\\
    $v_{resetTRN}$	& -60 mV	\\
    $v_{switch}$ & -70 mV \\
	$v_{peakTRN}$	& 25 mV	\\
    $a$ & $1$ nS \\
	$b$ & 50 pA	\\
    $k_{spikelet}$ &	40 \\
   \end{tabularx} \\ \\
 \end{tabular} 
 
 &
 
 \begin{tabular}[t]{l}
 \textbf{Thalamo-Cortical Neurons} \\
   \begin{tabularx}{.4\linewidth}{l r}
	$\tau_{E}$	& 12 ms	\\
	$\tau_m$	& 40 ms \\
	$R_{m}$	& 0.6 $\Omega$	\\
	$v_{resetTC}$	&	-70 mV	\\
  	$v_{threshTC}$ &	0 mV \\
   \end{tabularx}\\ \\
   
   \textbf{Gap junction plasticity} \\
  	\begin{tabularx}{.4\linewidth}{l r}
	$\alpha_{gLTD}$	& $15.69$ nS $\cdot$ ms$^{-1}$	\\
	$\alpha_{gLTP}$	&	15 $\alpha_{gLTD}$ \\
	$\theta_{q}$	&	0.3\\ 
	$\tau_{b}$	&	8 ms \\
	$\tau_{q}$	&	6000 ms \\
	\end{tabularx}\\ \\

  \end{tabular} \\ \\
	
  \begin{tabular}[t]{l}
  
   \textbf{Network} \\
   \begin{tabularx}{.4\linewidth}{l r}
    dt &	1 ms	\\
	$N_I$	&	100	\\
	$N_E$	&	200 \\
	$W^{II}$	& $-200$	\\
	$W^{IE}$	& $-1000$	\\
	$W^{EE}$	& $500$	\\
	$W^{EI}$	& $300$	\\
	$\gamma_b$ & 	13	\\ 
    $\sigma_{gap}$ & 1 \\
    $\mu_{gap}$ & 1 \\
    $\sigma_I$	&	400 pA\\
	$\nu_I$ &	[0 pA; 300 pA]	\\ 
	$\tau_{noise}$ & 10 ms \\
	\end{tabularx} 
\end{tabular} &

\begin{tabular}[t]{l}
 \textbf{Spindle analysis} \\
   \begin{tabularx}{.4\linewidth}{l r}
	$NFFT$	& 2000 ms	\\
	$\theta_{spindle}$	&	0.5 dB	\\
   \end{tabularx}
   \end{tabular}

\end{tabular}
 
\label{table:params1} 
\end{table}

\newpage
\bibliography{library}

\newpage

%
%

\newpage
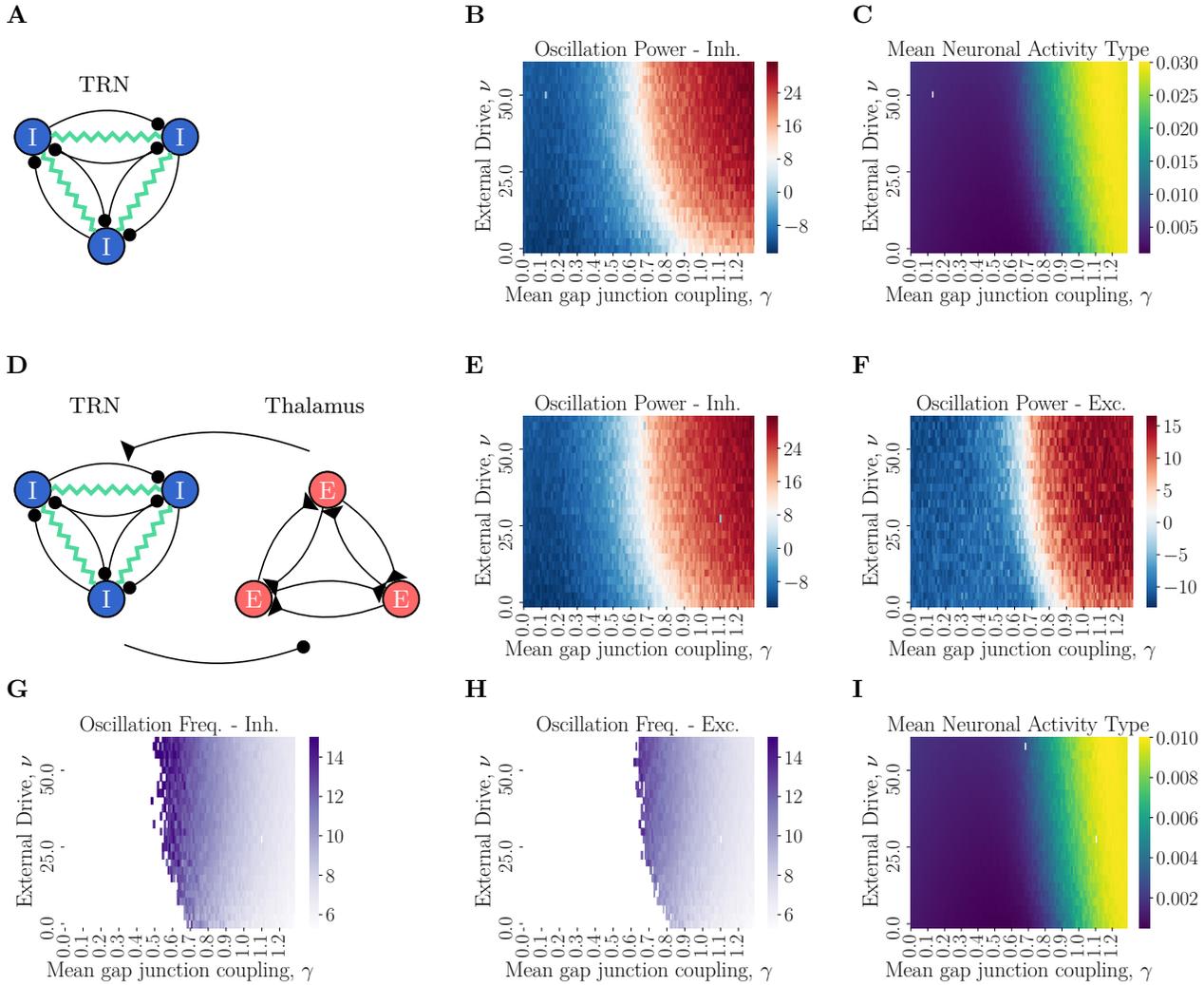
\begin{figure}[H]
\begin{fullwidth}
	\begin{tabular}{ p{14cm}}
		\begin{tabular}{p{6cm} p{5cm} p{5cm}}
		
			\textbf{\large{A}} &  \textbf{\large{B}} & \textbf{\large{C}} \\
			\usetikzlibrary{matrix,chains,positioning,decorations.pathreplacing,arrows,arrows.meta}
\usetikzlibrary{matrix,positioning,calc}
\usetikzlibrary{decorations.pathmorphing}

\xdefinecolor{c1}{HTML}{4ED99C}
\xdefinecolor{Ecol}{HTML}{FF6868}
\xdefinecolor{Icol}{HTML}{3366cc}

\begin{tikzpicture}[
           > = Stealth, semithick, 
plain/.style = {draw=none, fill=none, yshift=11mm,
                text width=7ex,  align=center},
   ec/.style = {draw=none},
  net/.style = {
    matrix of nodes,
    nodes={circle, draw, semithick, minimum size=5.1mm, inner sep=0mm},
    nodes in empty cells,
  column sep = 5mm, 
     row sep = 0mm  
            },
            net2/.style = {
    matrix of nodes,
     nodes={ minimum size=5.1mm, inner sep=0mm},
    nodes in empty cells,
  column sep = 10mm, 
     row sep = 0mm  
            },
]

\matrix[net2] (m2) at  (0,1.5)
{
  TRN  & |[ec]|      & |[ec]|     \\ [0pt]
};

\matrix[net] (m) at  (0,0)
{
|[ec]|                    & |[ec]|                  & |[ec]|    \\ [0pt]
   I                   & |[ec]|                  & I     \\ [9mm]
|[ec]|                &     I                    &  |[ec]|    &   [-2pt]
|[ec]|                    & |[ec]|                  & |[ec]|   \\ 
|[ec]|                    & |[ec]|                  & |[ec]|   \\ 
};
\% inputs

\tikzset{I/.style args={[#1]#2}{
    draw,
    circle,
    fill=Icol,
    minimum size=5mm,
    label={[white,#1]center:#2}
    }
}

\node at (m-2-1) [I={[font=\large]I}] {};
\node at (m-2-3) [I={[font=\large]I}] {};
\node at (m-3-2) [I={[font=\large]I}] {};

\path[-*]
	(m-2-1) edge [bend left=30,looseness=1] (m-2-3)
			edge [bend left=30,looseness=1] (m-3-2)
	(m-2-3) edge [bend left=30,looseness=1] (m-3-2)
			edge [bend left=30,looseness=1] (m-2-1)
	(m-3-2) edge [bend left=30,looseness=1] (m-2-1)
			edge [bend left=30,looseness=1] (m-2-3);

\path
	(m-2-1) edge [decoration = {zigzag,segment length = 2mm, amplitude = 0.5mm},decorate,ultra thick,c1] (m-2-3)
	(m-2-3) edge [decoration = {zigzag,segment length = 2mm, amplitude = 0.5mm},decorate,ultra thick,c1] (m-3-2)
	(m-3-2) edge [decoration = {zigzag,segment length = 2mm, amplitude = 0.5mm},decorate,ultra thick,c1] (m-2-1);

\end{tikzpicture}
			&
			\resizebox{5cm}{!}{\includegraphics{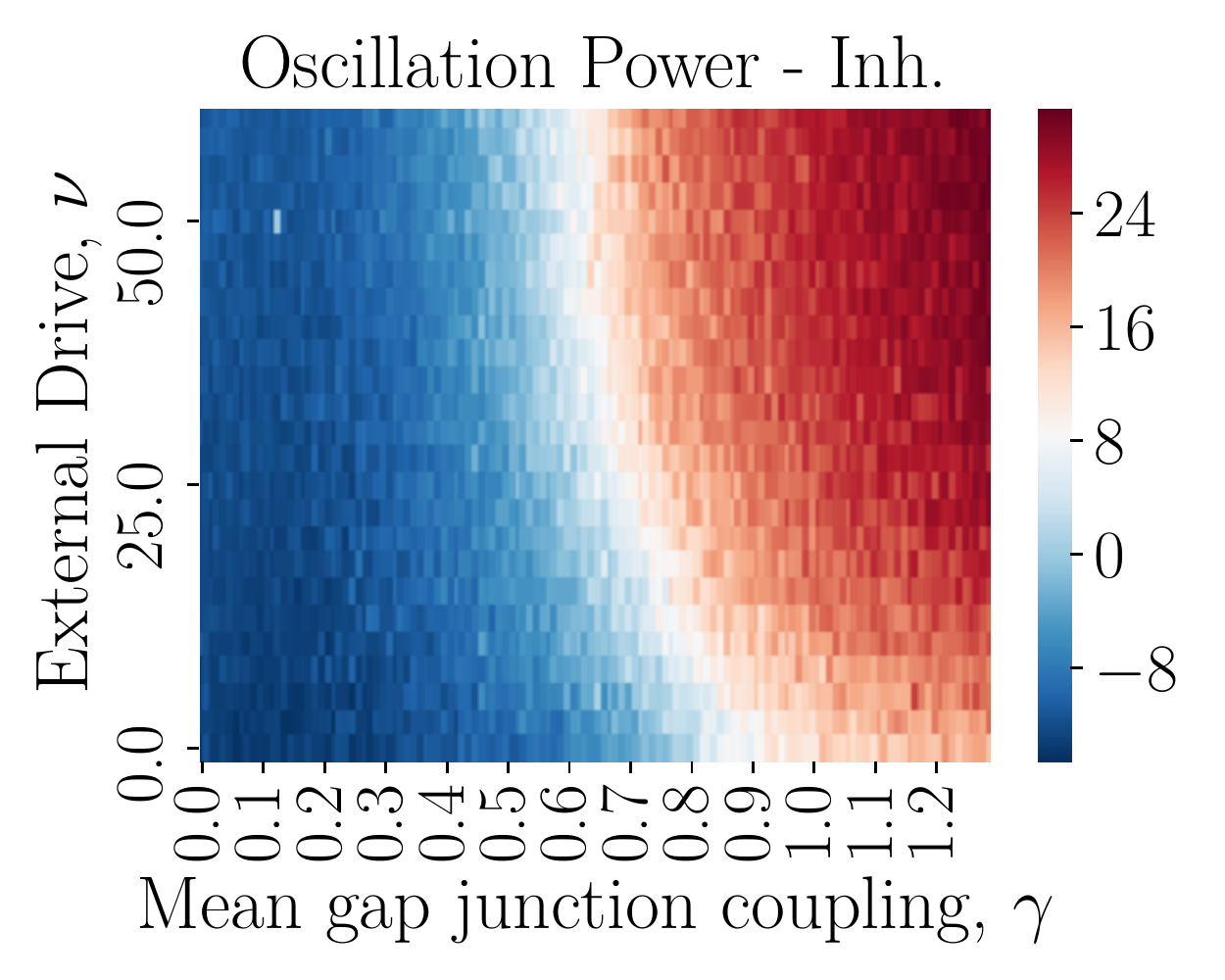}}
			&
			\resizebox{5cm}{!}{\includegraphics{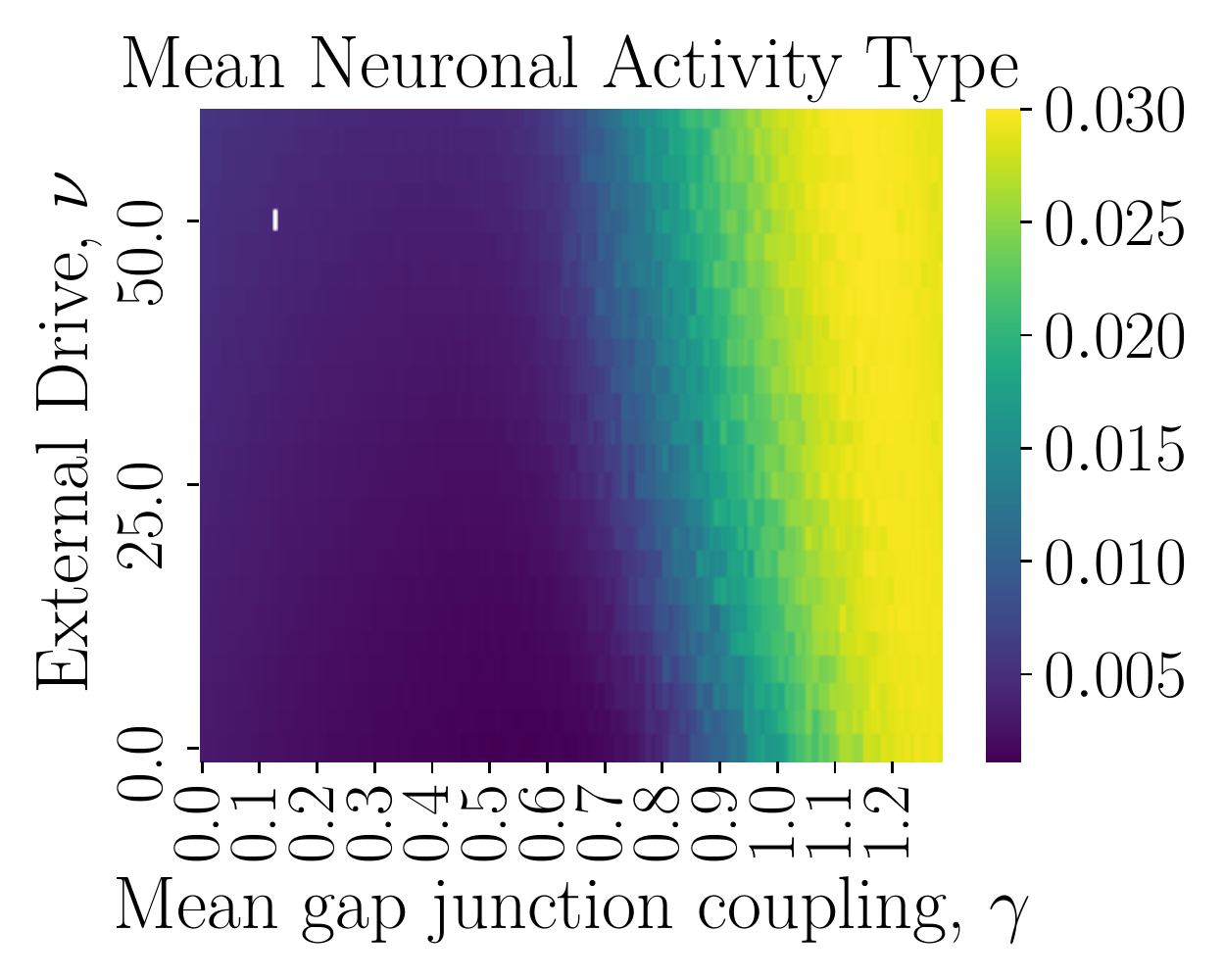}}

		\end{tabular}

		\begin{tabular}{p{6cm} p{5cm} p{5cm}}
		
			\textbf{\large{D}} &  \textbf{\large{E}} & \textbf{\large{F}} \\
			\begin{tikzpicture}[
           > = Stealth, semithick, 
plain/.style = {draw=none, fill=none, yshift=11mm,
                text width=7ex,  align=center},
   ec/.style = {draw=none},
  net/.style = {
    matrix of nodes,
    nodes={circle, draw, semithick, minimum size=5.1mm, inner sep=0mm},
    nodes in empty cells,
  column sep = 5mm, 
     row sep = 0mm  
            },
            net2/.style = {
    matrix of nodes,
    nodes in empty cells,
  column sep = 18mm, 
     row sep = 0mm  
            },
]

\matrix[net2] (m2) at  (0,0)
{
 TRN       &  Thalamus      \\ [-2pt]
};

\matrix[net] (m) at  (0,-2)
{
|[ec]|                    & |[ec]|                  & |[ec]|     &    |[ec]| &  |[ec]|     &   |[ec]| \\ [-2pt]
   I                   & |[ec]|                  & I    &   |[ec]|   &  E    &   |[ec]| \\ [9mm]
|[ec]|                &     I                    &  |[ec]|    &     E                   &  |[ec]|  & E        \\ [-2pt]
|[ec]|                    & |[ec]|                  & |[ec]|     & |[ec]| &  |[ec]|     &   |[ec]| \\ 
};
\% inputs

%

\tikzset{E/.style args={[#1]#2}{
    draw,
    circle,
    fill=Ecol,
    minimum size=5mm,
    label={[white,#1]center:#2}
    }
}
\tikzset{I/.style args={[#1]#2}{
    draw,
    circle,
    fill=Icol,
    minimum size=5mm,
    label={[white,#1]center:#2}
    }
}

\node at (m-2-5) [E={[font=\large]E}] {};
\node at (m-3-4) [E={[font=\large]E}] {};
\node at (m-3-6) [E={[font=\large]E}] {}; 

\node at (m-2-1) [I={[font=\large]I}] {};
\node at (m-2-3) [I={[font=\large]I}] {};
\node at (m-3-2) [I={[font=\large]I}] {};

\path[-*]
	(m-2-1) edge [bend left=30,looseness=1] (m-2-3)
			edge [bend left=30,looseness=1] (m-3-2)
	(m-2-3) edge [bend left=30,looseness=1] (m-3-2)
			edge [bend left=30,looseness=1] (m-2-1)
	(m-3-2) edge [bend left=30,looseness=1] (m-2-1)
			edge [bend left=30,looseness=1] (m-2-3);

\path[-triangle 90 reversed]
	(m-2-5) edge [bend left=20,looseness=1] (m-3-4)
			edge [bend left=20,looseness=1] (m-3-6)
	(m-3-4) edge [bend left=20,looseness=1] (m-2-5)
			edge [bend left=20,looseness=1] (m-3-6)
	(m-3-6) edge [bend left=20,looseness=1] (m-2-5)
			edge [bend left=20,looseness=1] (m-3-4);
			
\path[-*]
	(m-4-2) edge [bend right=20,looseness=1] (m-4-5);
	
\path[-triangle 90 reversed]
	(m-1-5) edge [bend right=20,looseness=1] (m-1-2);
			
\path
	(m-2-1) edge [decoration = {zigzag,segment length = 2mm, amplitude = 0.5mm},decorate,ultra thick,c1] (m-2-3)
	(m-2-3) edge [decoration = {zigzag,segment length = 2mm, amplitude = 0.5mm},decorate,ultra thick,c1] (m-3-2)
	(m-3-2) edge [decoration = {zigzag,segment length = 2mm, amplitude = 0.5mm},decorate,ultra thick,c1] (m-2-1);

\end{tikzpicture}
			&
			\resizebox{5cm}{!}{\includegraphics{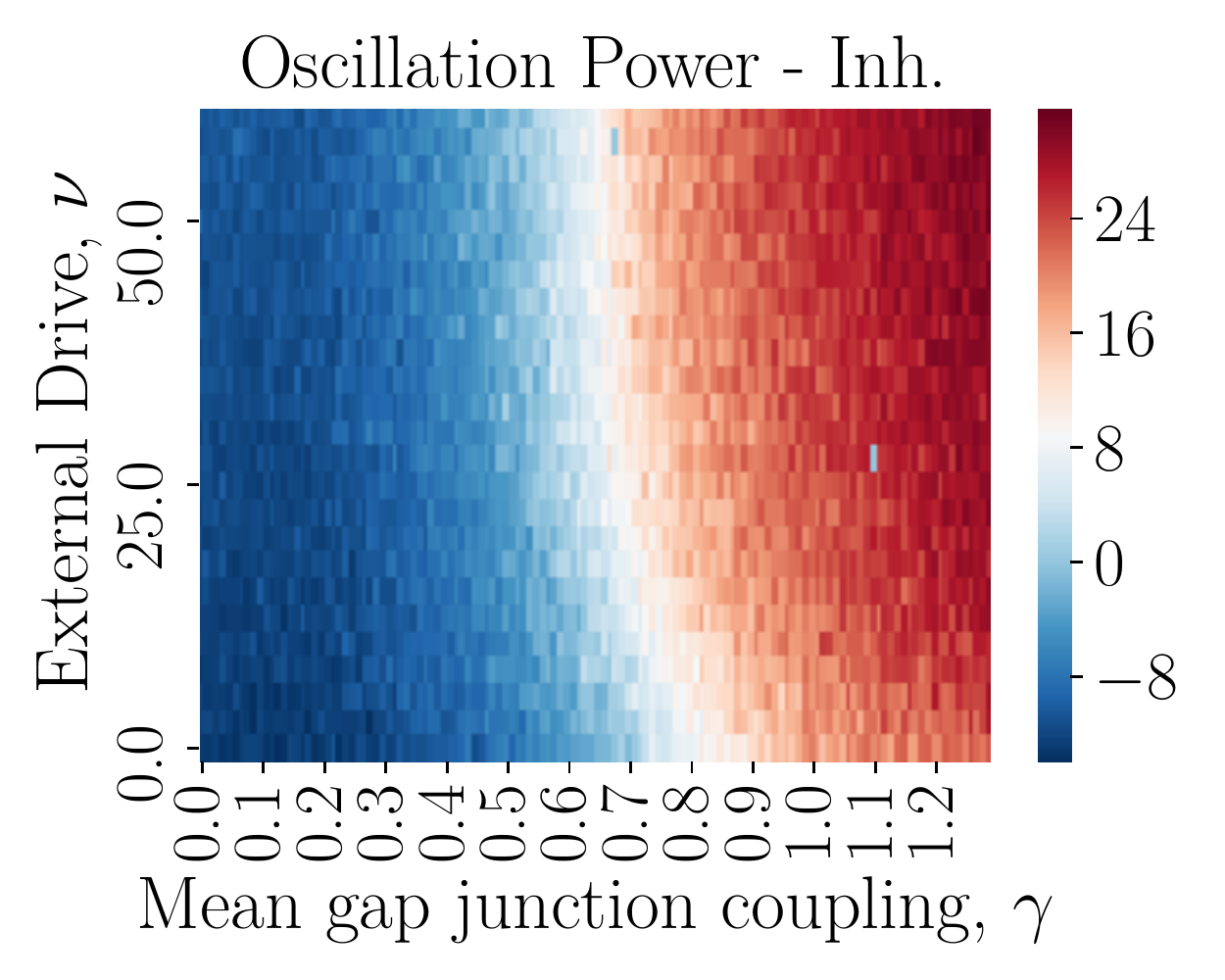}}
			&
			\resizebox{5cm}{!}{\includegraphics{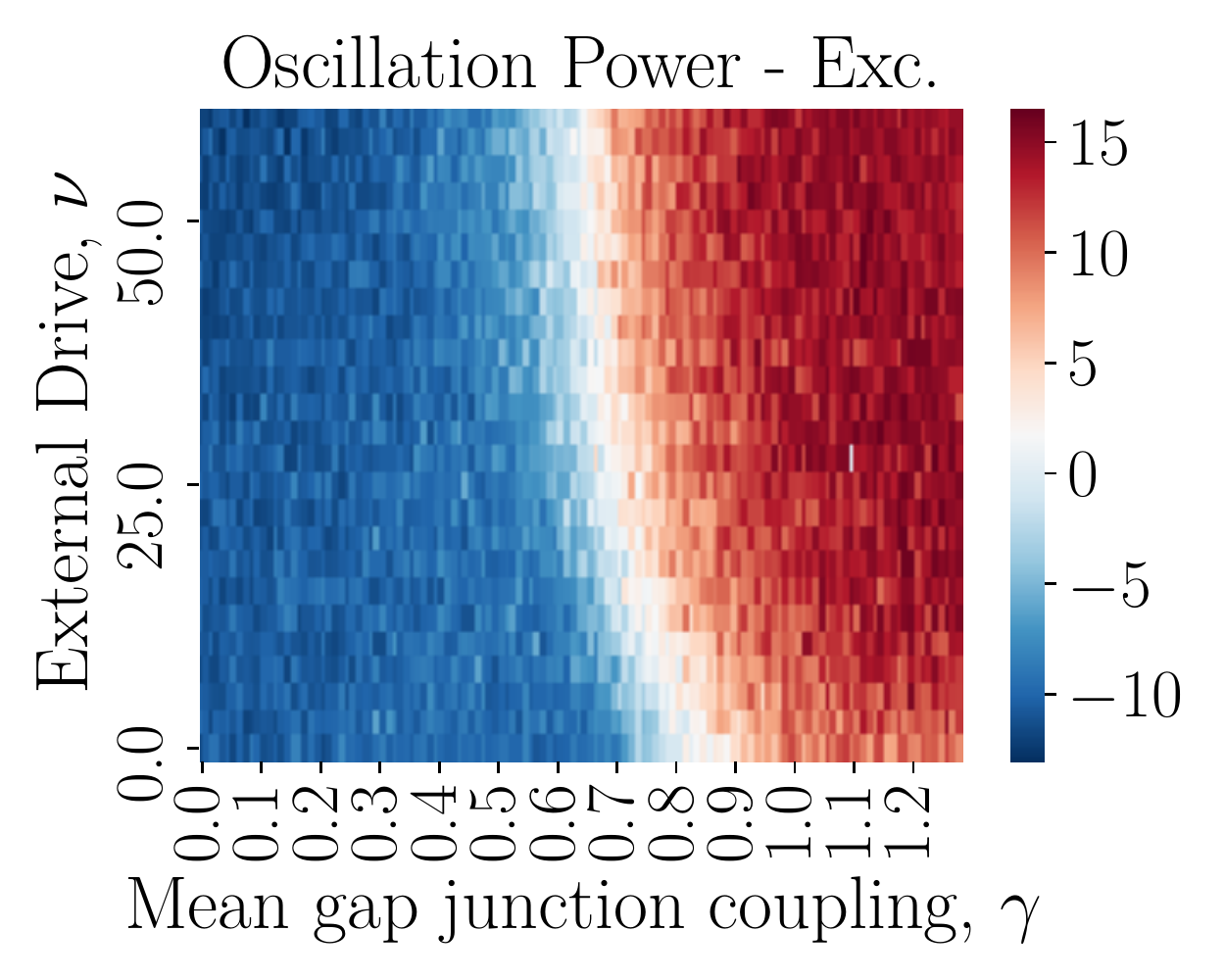}}
			\\
			\textbf{\large{G}} &  \textbf{\large{H}} & \textbf{\large{I}} \\
			\resizebox{5cm}{!}{\includegraphics{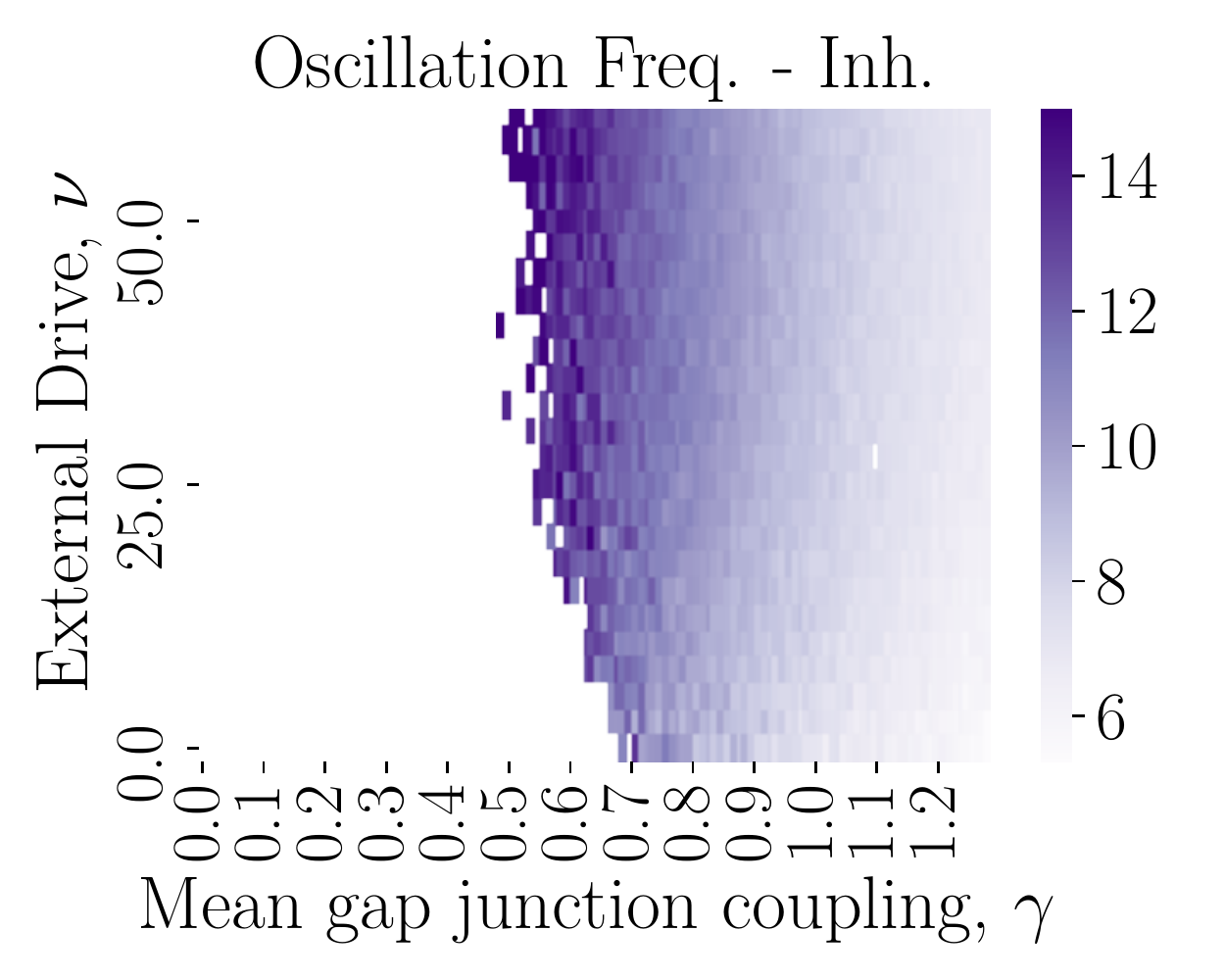}}
			&
			\resizebox{5cm}{!}{\includegraphics{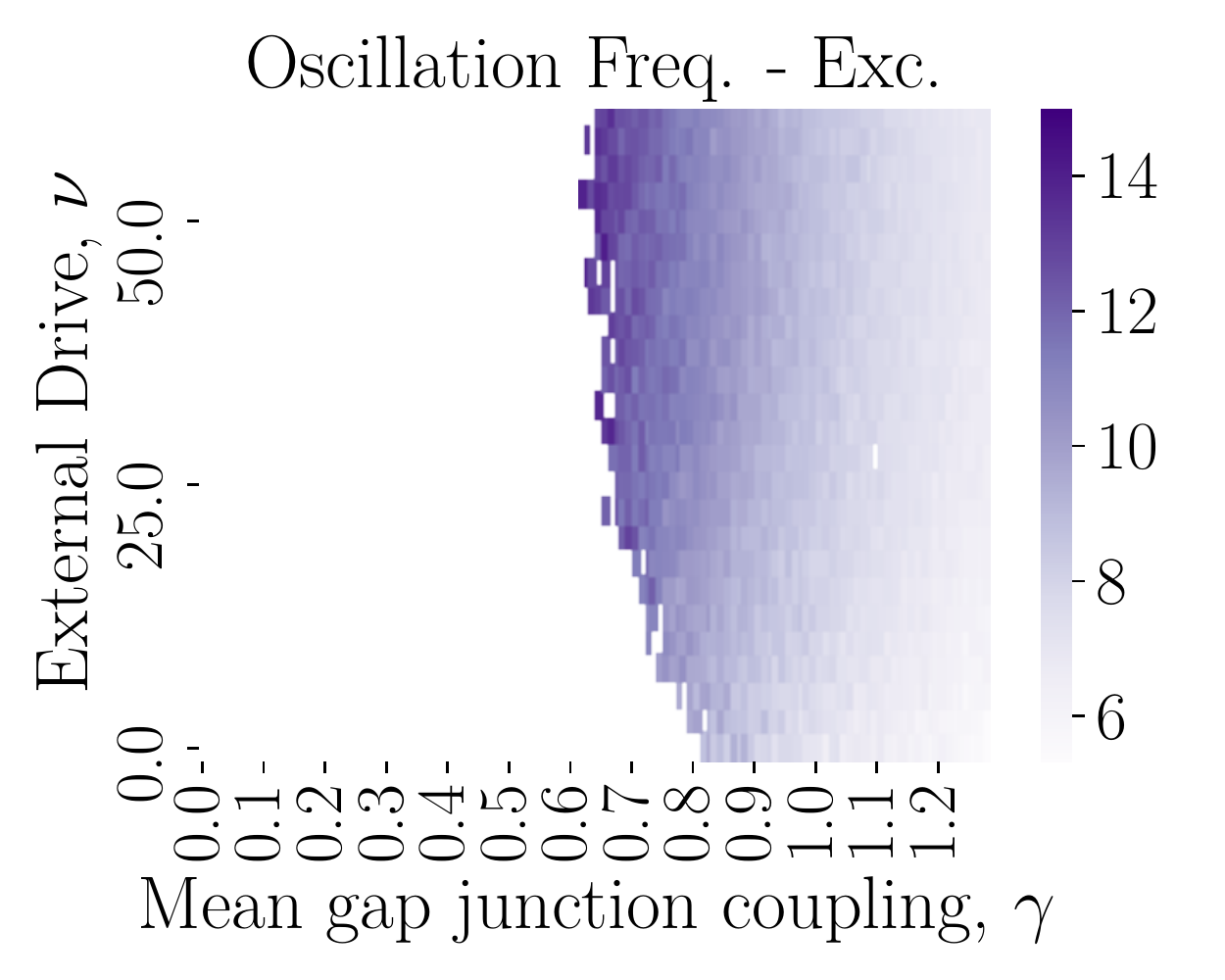}}
			&
			\resizebox{5cm}{!}{\includegraphics{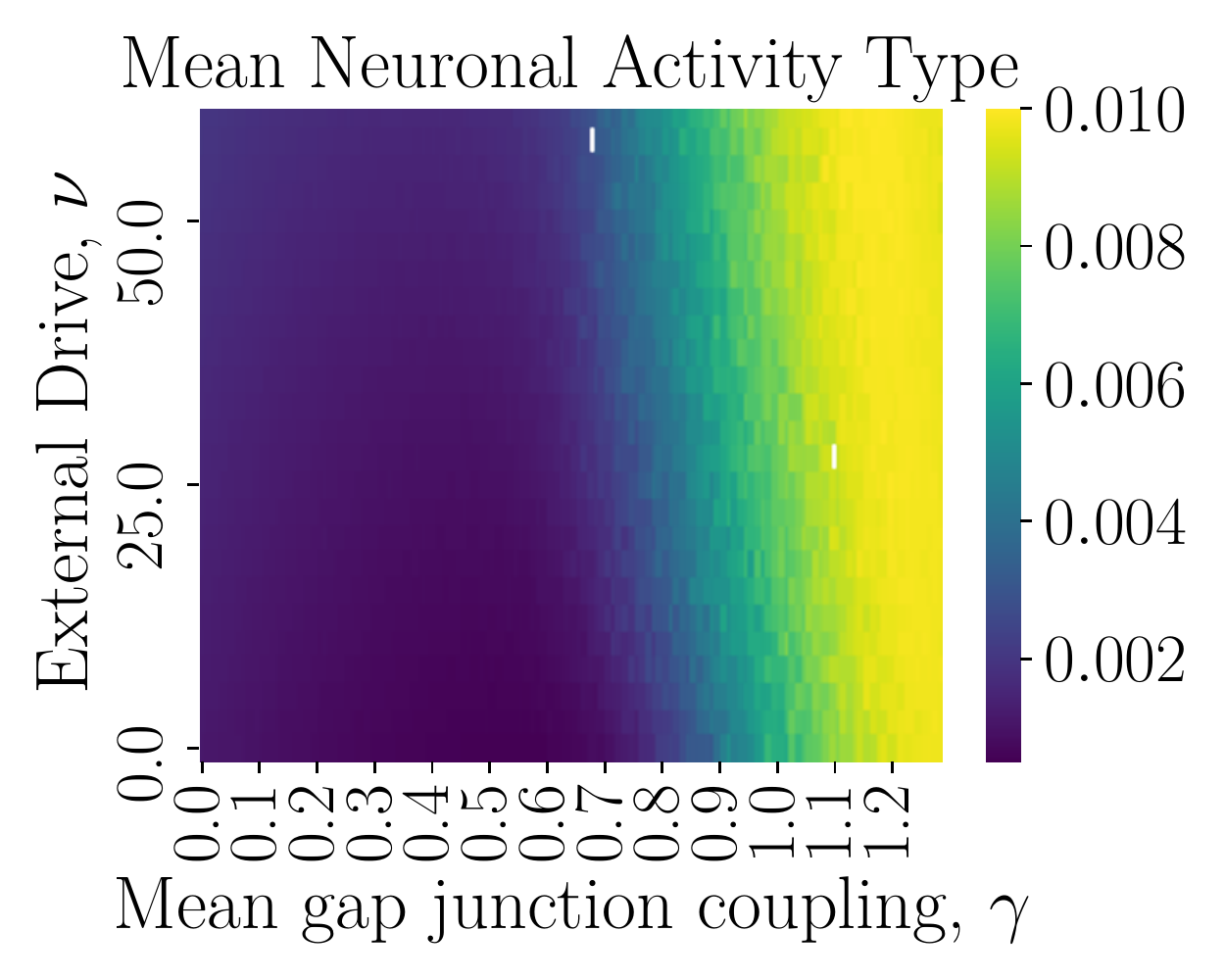}}

		\end{tabular}

	\end{tabular}
	\caption{ {\bf TRN-TC synchrony depends on gap junction strength in TRN.} \label{fig1}}
	(\textbf{A}) The TRN network consists of inhibitory (I) neurons coupled in an all-to-all fashion with chemical synapses (black lines) and gap junctions (jagged green lines).
	For all the following activity diagrams, the x-axis represents the mean gap junction coupling strength $\gamma$ and the y-axis represents the mean value of the network external drive $\nu$.
	(\textbf{B}) Power in the Fourier domain of the strongest oscillation frequency of the population activity. The red area denotes oscillations characterizing the synchronous regular regime (SR) and the blue area show the lack of oscillations characterizing the asynchronous irregular regime (AI).
	(\textbf{C}) Ratio of bursting $A_{bursting}$ over spiking $A_{spiking}$ activity. Sparse spiking prevails in the dark region while bursting is dominating in the light region.
	(\textbf{D}) The thalamus network consists of excitatory neurons coupled with chemical synapses among them and with the TRN inhibitory neurons in an all-to-all fashion.
	(\textbf{E}) Same as \textbf{B}, but for the TRN inhibitory neurons of the TRN-TC network.
	(\textbf{F}) Same as \textbf{B}, but for the TC excitatory neurons of the TRN-TC network.
	(\textbf{G}) Strongest oscillation frequency of the population activity of the TRN inhibitory neurons. Most values are contained in the 7 - 15 Hz spindle frequency range.
	(\textbf{H}) Same as \textbf{G}, but for the TC excitatory neurons. 
	(\textbf{I}) Same as \textbf{C}, but for the TRN inhibitory neurons of the TRN-TC network.

\end{fullwidth}
\end{figure}
 
%
%
\newpage
\begin{figure}[H]
\begin{fullwidth}
	\begin{tabular}{ p{18cm}}
		\begin{tabular}{ p{18cm}}
	        		\textbf{\large{A}} \\
	        		\resizebox{12cm}{!}{\includegraphics{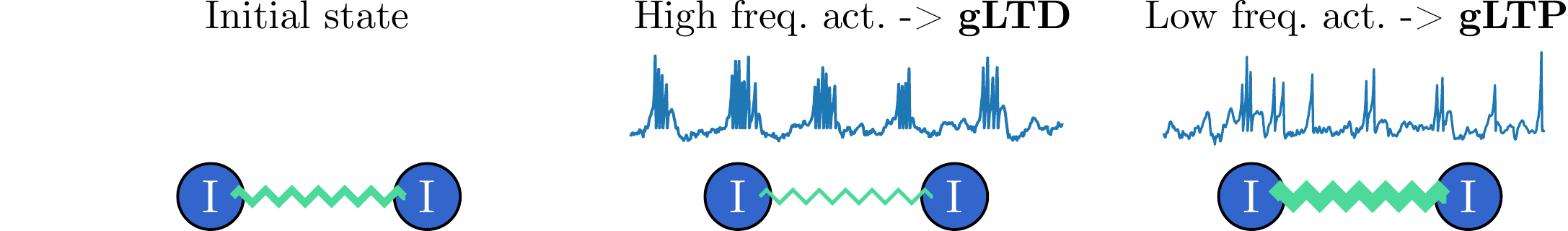}}
	        		\\
	        		\textbf{\large{B}} \\
				\resizebox{18cm}{!}{\includegraphics{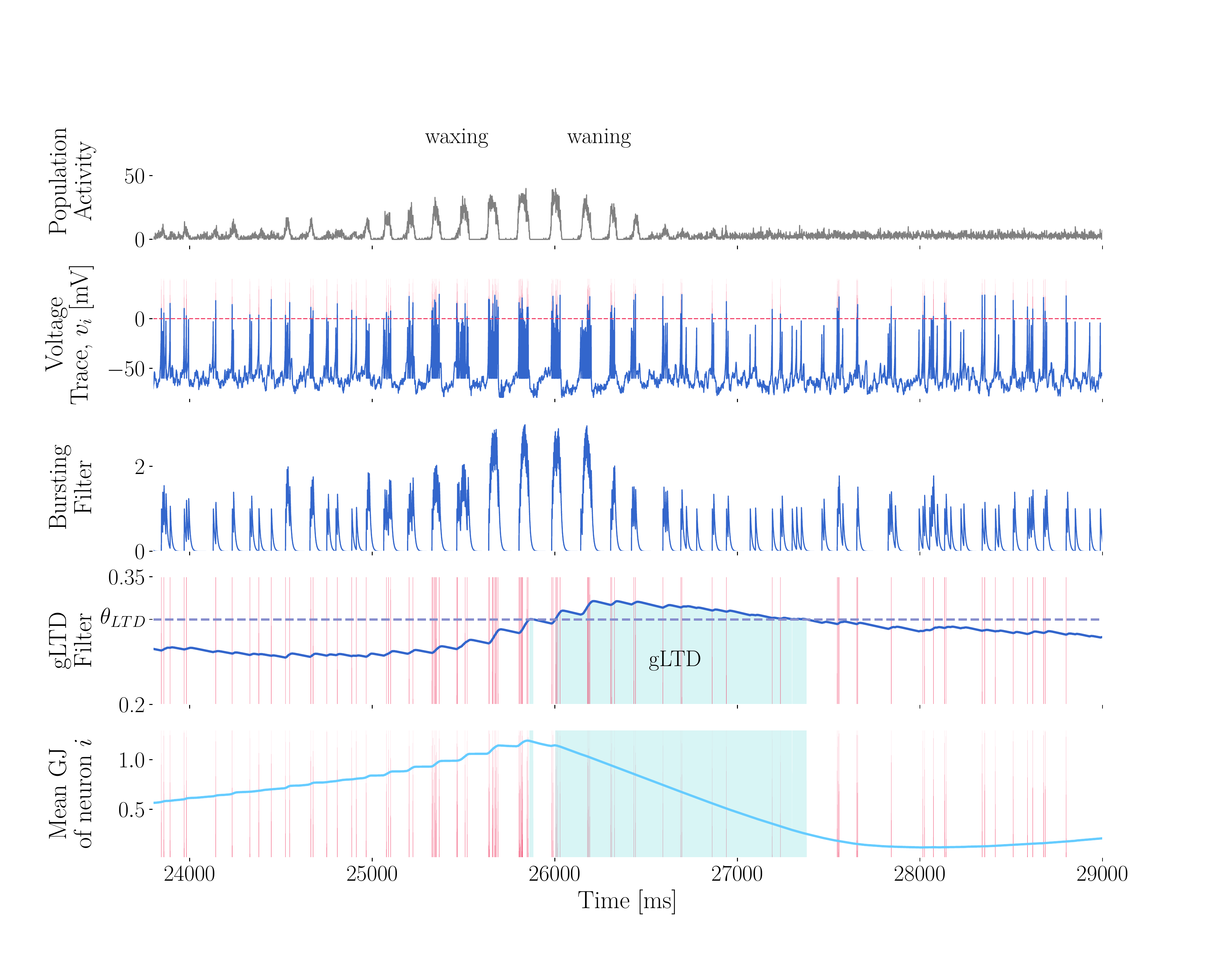}}
		\end{tabular}
	\end{tabular}
	\\
	\caption{ {\bf Gap junction plasticity can generate spindle-like activity.} }
	(\textbf{A}) The gap junction plasticity is activity-dependent, high-frequency activity leads to long-term depression of the gap junctions (gLTD), while sparse activity leads to long-term potentiation of the gap junctions (gLTP).
	(\textbf{B}) Plasticity mechanism leading to the generation of spindle-like events. The first row shows the population activity of inhibitory neurons, defined as the sum of action potentials, as a function of time. After a period of asynchronous activity, neurons synchronize in a waxing and waning fashion. 
	The second row shows the voltage trace of a single neuron $i$. The same neuron $i$ is chosen for the following graphs. The vertical red bars show the neuron spiking times ($v_i \geq 0$). 
	The third row shows the low-pass filter of the spiking activity, quantifying the bursting activity of the neuron.
	The fourth row shows the second low-pass filter of the spiking activity, which measures sustained bursting activity. The dashed-line ($\theta_{LTD}$ = 0.3) represents the threshold above which gLTD (green region) is activated.
	(\textbf{D}) The last row shows the mean value of all gap-junctions coupling neuron $i$. The gap junctions are potentiated on the spike times (vertical red bars) and depressed when the LTD filter is above its threshold.

    \end{fullwidth}
\end{figure}

\newpage
\begin{figure}[H]
\begin{fullwidth}
	\begin{tabular}{p{19cm}}
		\begin{tabular}{p{19cm}}
			\textbf{\large{A}}\\			
			\resizebox{19cm}{!}{\includegraphics{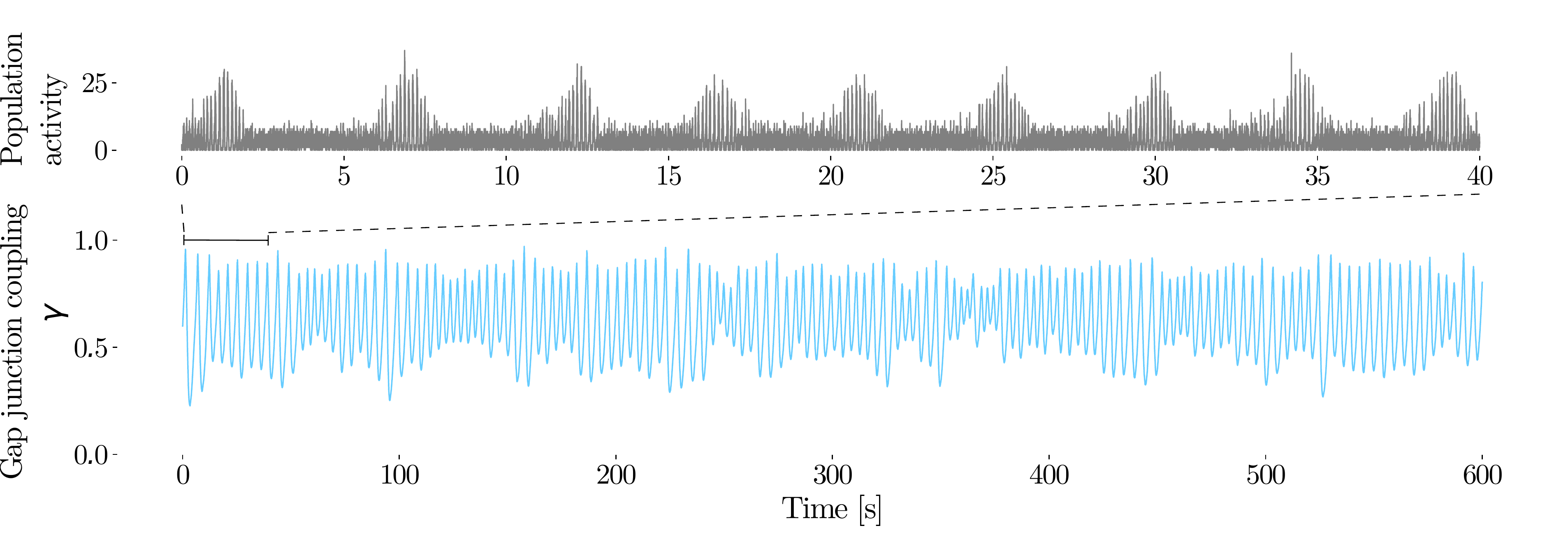}}\\
	
		\end{tabular}

	\end{tabular}

	\begin{tabular}{ p{20cm}}
		\begin{tabular}{  p{4cm} p{8cm}}
	        		\textbf{\large{B}} & \textbf{\large{}} \\
		
				\resizebox{4cm}{!}{\includegraphics{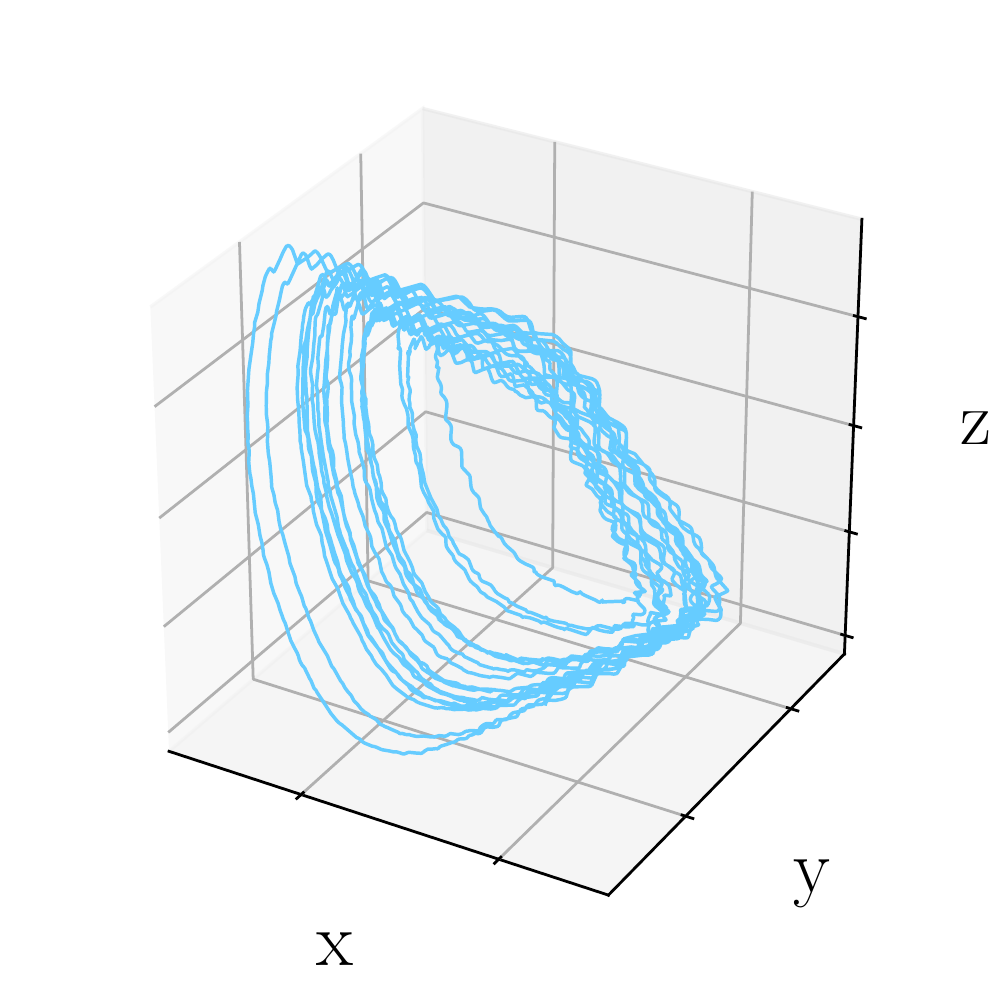}}
				&
				\resizebox{8cm}{!}{\includegraphics{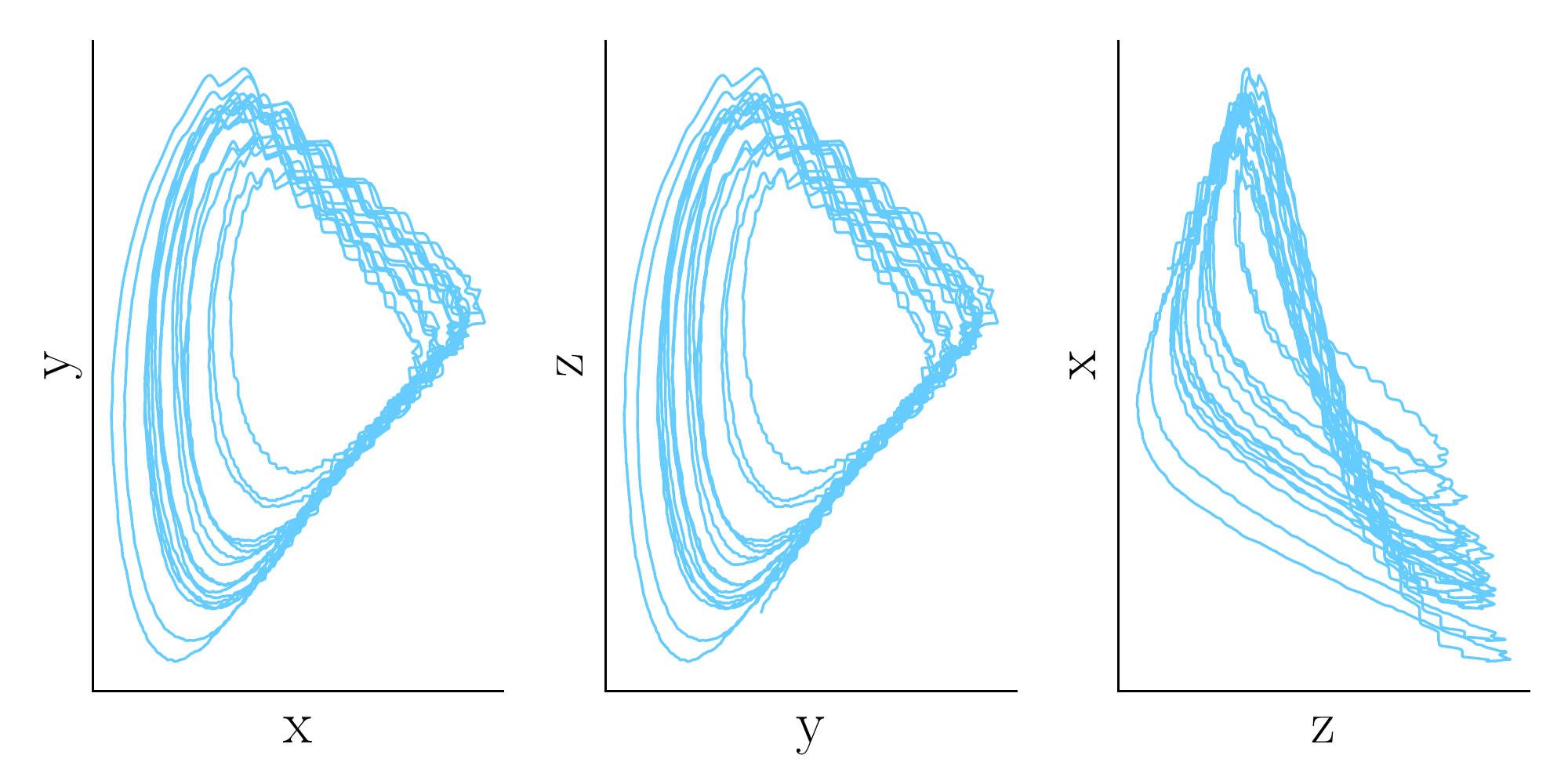}}
				
		\end{tabular}
	\end{tabular}
	
	\begin{tabular}{p{20cm}}
		\begin{tabular}{p{12cm}p{6cm}}
			\textbf{\large{C}} & \textbf{\large{D}} \\
			\resizebox{12cm}{!}{\includegraphics{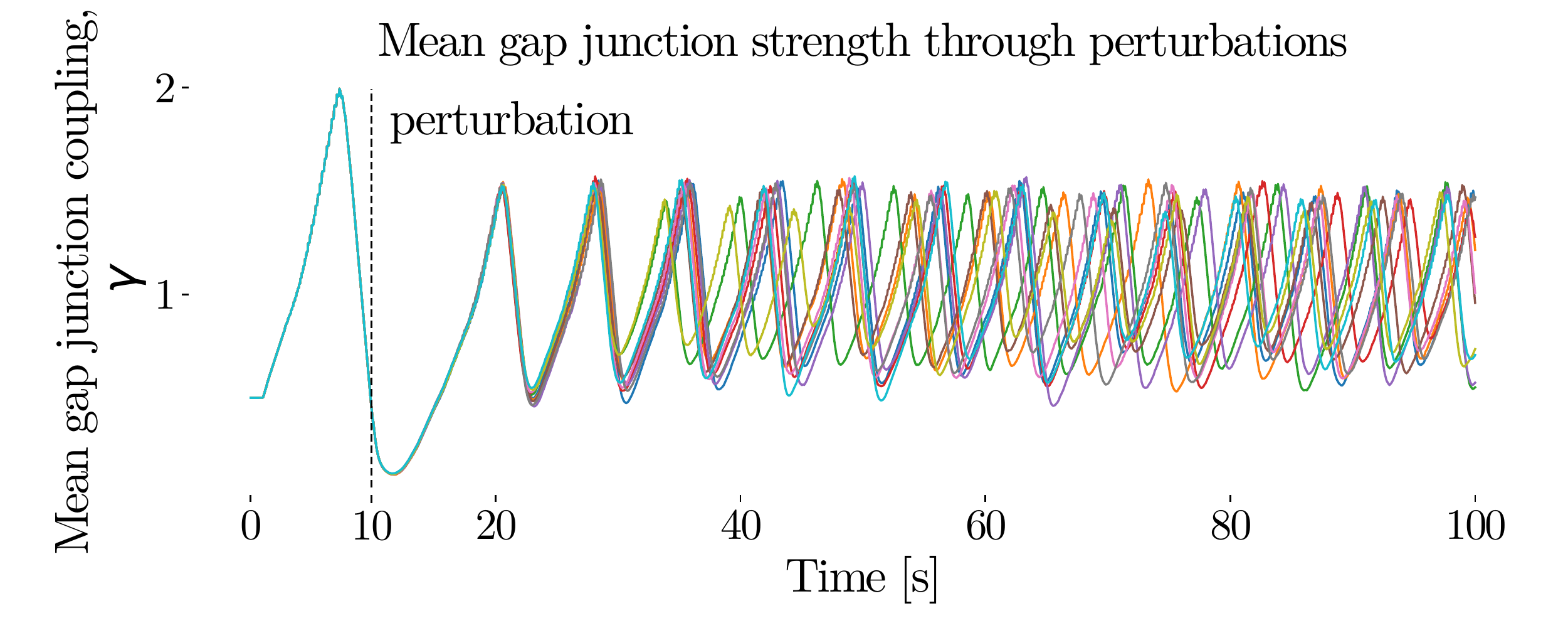}}
			&
				\resizebox{6cm}{!}{\includegraphics{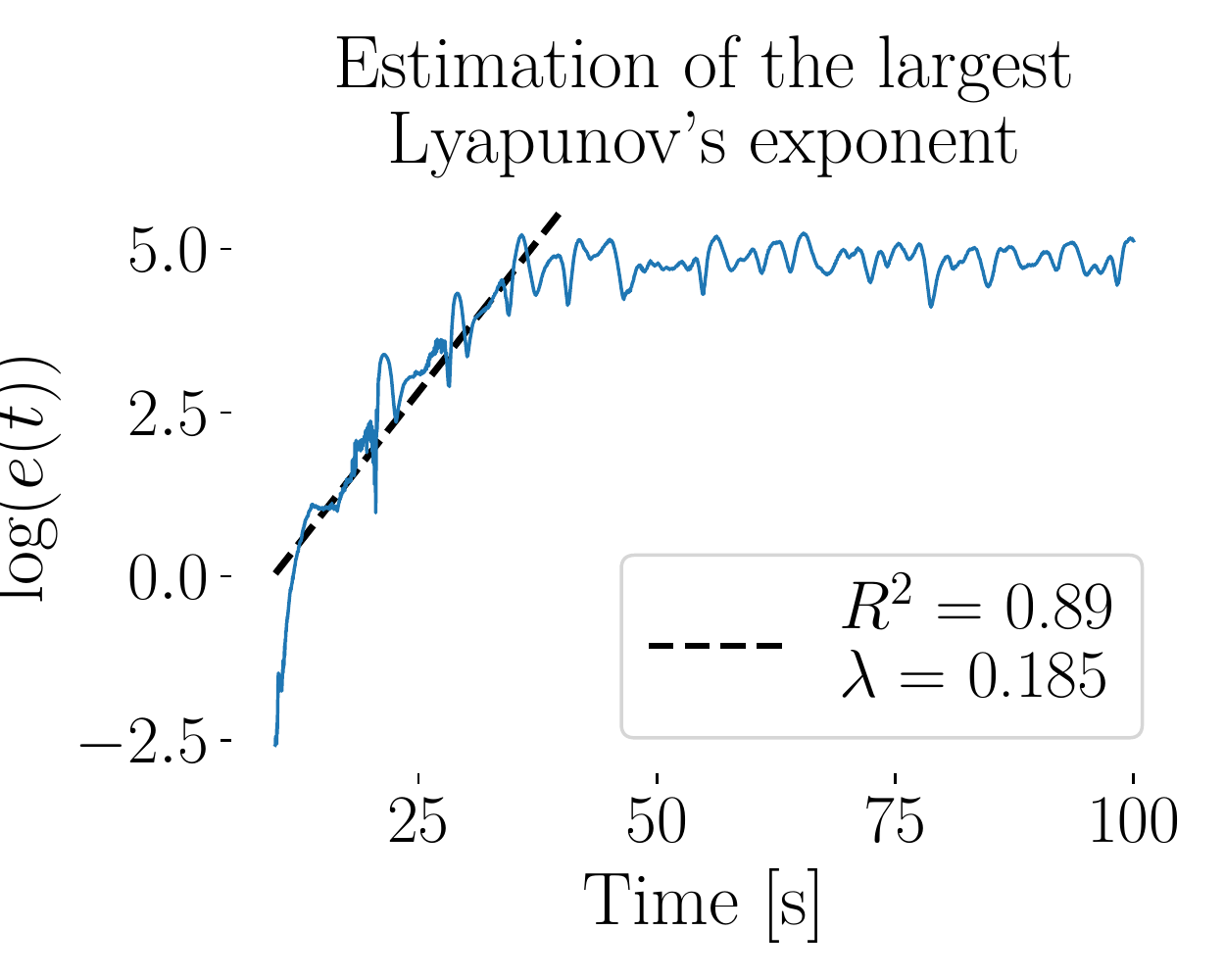}}  
		\end{tabular}
	\end{tabular}
	\\
	\caption{ {\bf Gap junction plasticity induces irregular behavior.} }
	(\textbf{A}) Population activity, defined at the instant sum of action potentials of inhibitory neurons (top, 40 seconds) and mean gap junction coupling (bottom, 600 seconds) as a function of time. 
	(\textbf{B}) 3D Phase portrait of the mean gap junction strength (x) as function of its delayed representation (y delayed by 100 ms and z delayed by 200 ms). 2D Phase portrait in (x,y), (y,z) and (z,x) phase space representation.
	(\textbf{C}) 10 realizations of the mean gap junction coupling. The same random seed is chosen for all 10 realizations, however at 10 s, a small random perturbation (with a different random seed between realizations) is added to the gap junction coupling, and the traces diverge. All simulations are done with the same frozen noise and the gap junction perturbations are the only source of variability between realizations.
	(\textbf{D}) Estimation of the largest Lyapunov's exponent. The blue line represents the logarithm of the average of the L1 distance between pairs of realizations. The realizations diverge exponentially immediately after the perturbations. The dashed black line represents the results of a linear regression for the 30 seconds following the perturbations.

    \end{fullwidth}
\end{figure}

%
%

\newpage
\begin{figure}[H]
\begin{fullwidth}
	\begin{tabular}{ p{16cm}}
		\begin{tabular}{ p{8cm} p{8cm}}
	        		\textbf{\large{A}} & \textbf{\large{B}} \\
				\resizebox{8cm}{!}{\includegraphics{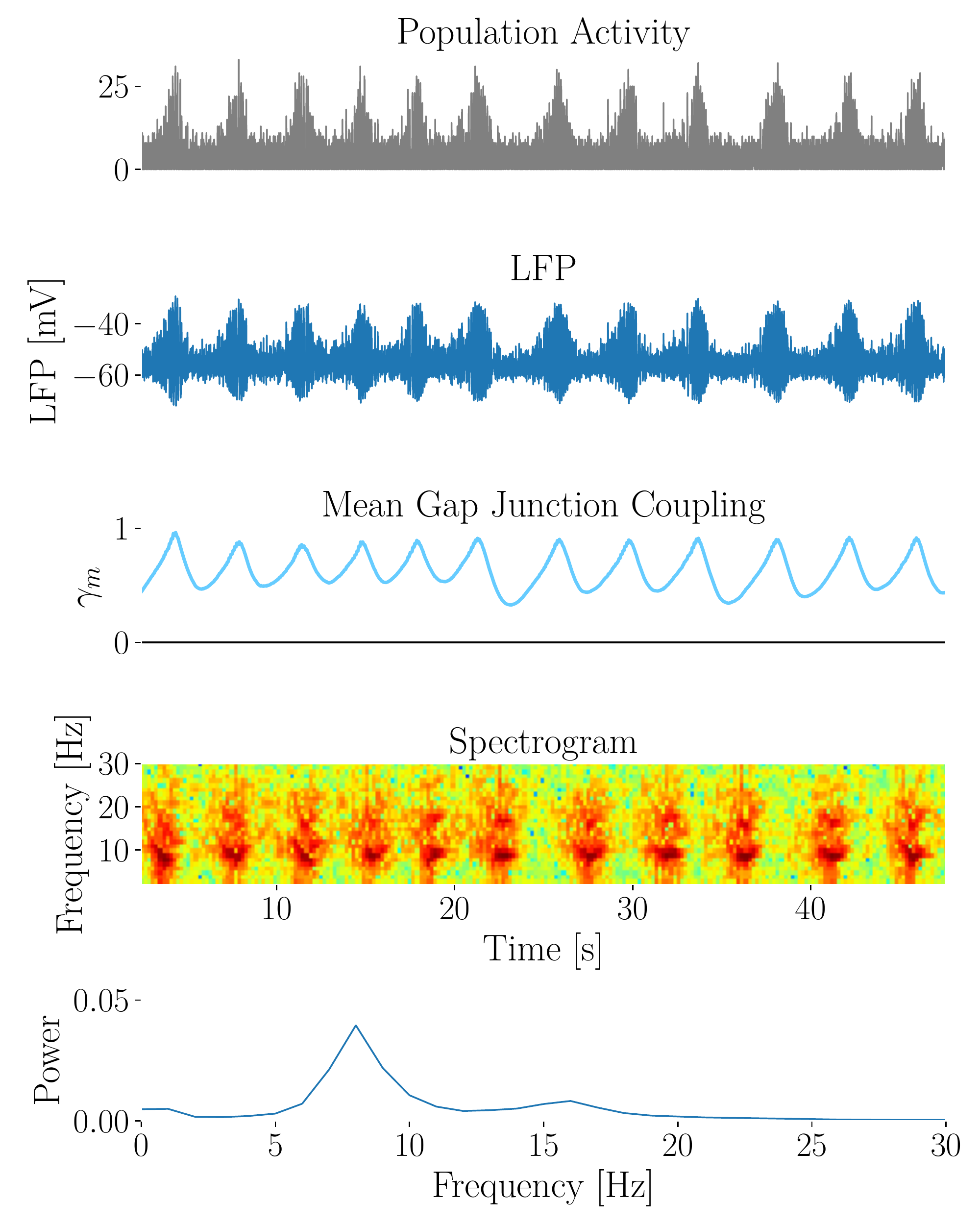}}
				&
				\resizebox{8cm}{!}{\includegraphics{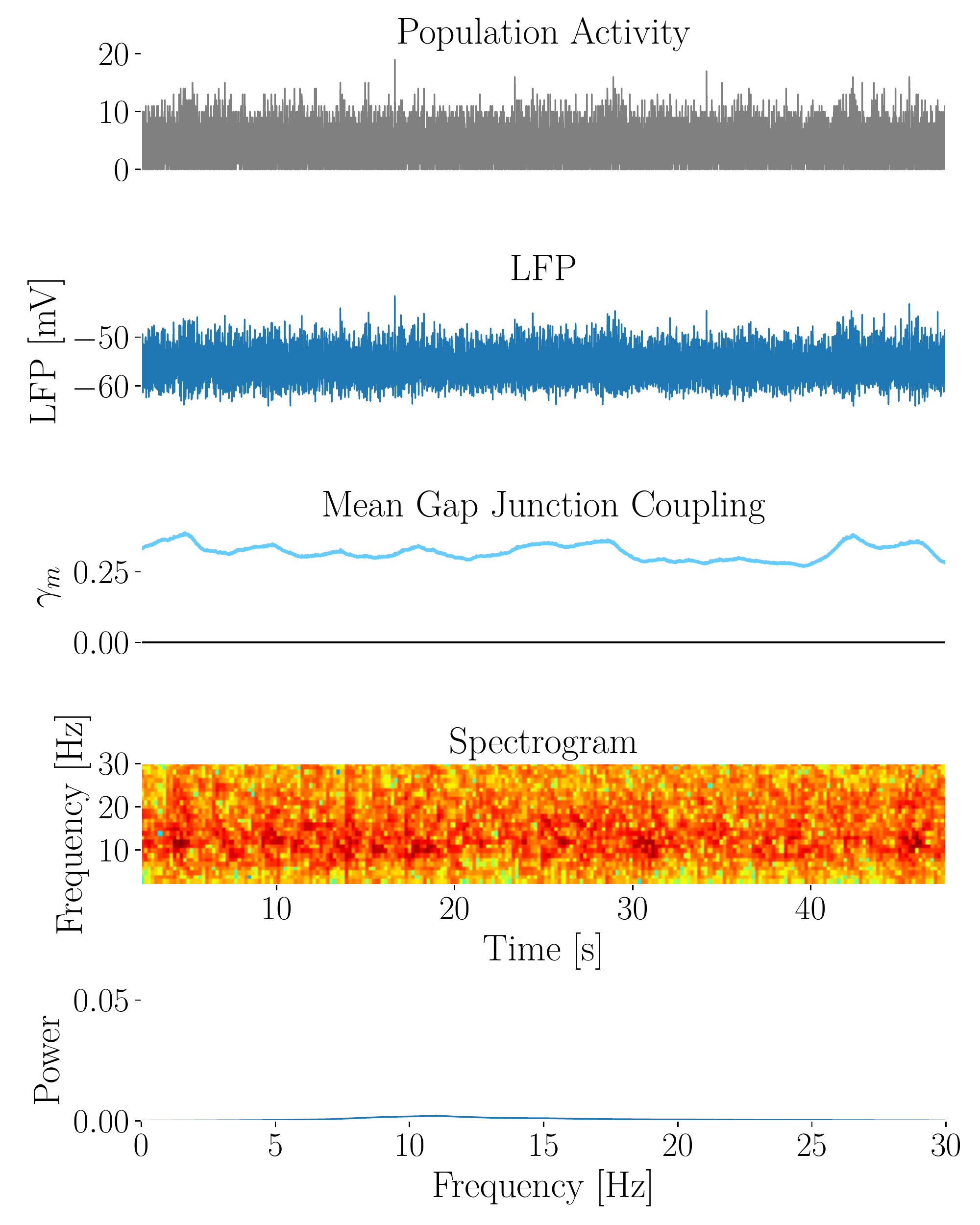}}
		\end{tabular}
		
		\begin{tabular}{ p{8cm} p{8cm}}
	        		\textbf{\large{C}} & \textbf{\large{D}} \\
				\resizebox{6cm}{!}{\includegraphics{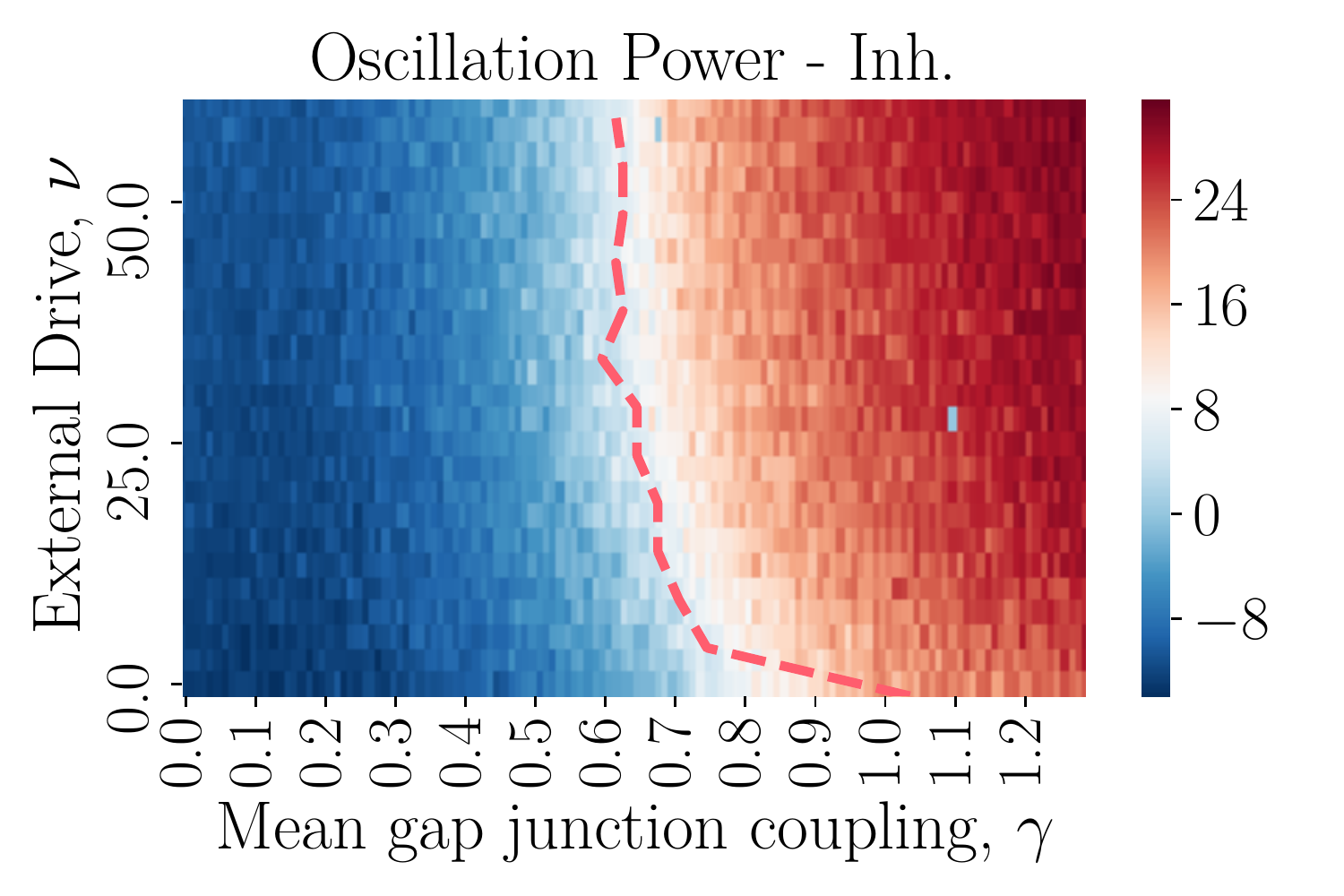}}
	            		&
				\resizebox{6cm}{!}{\includegraphics{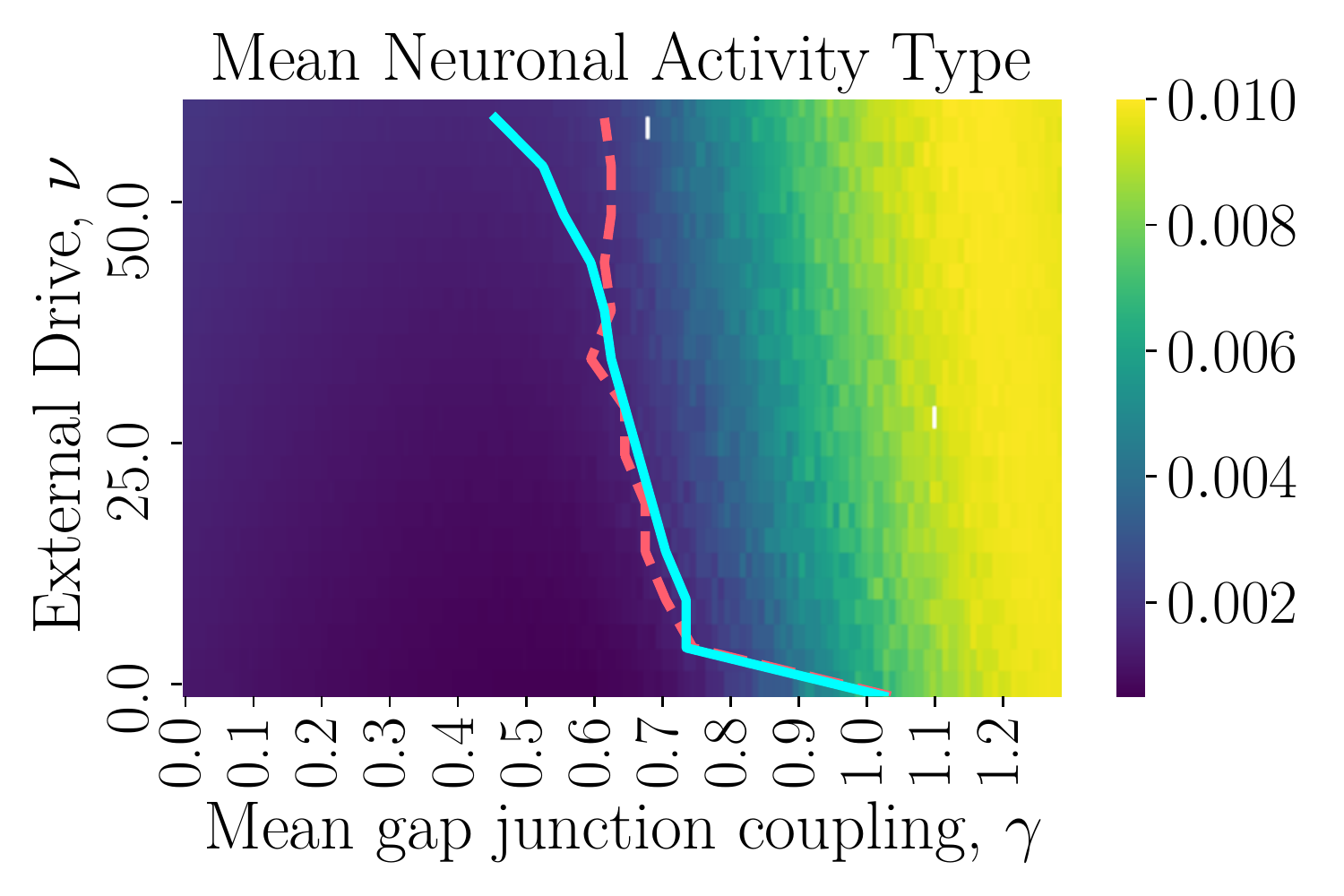}}

		\end{tabular}
	\end{tabular}
	\\
	\caption{ {\bf TRN activation can suppress spindle activity.} }
	Panels A and B consist of the same sub-panels, but for different values of TRN external drive (left: $\nu$ = 40 pA and right: $\nu$ = 50 pA). The first row shows the population activity of inhibitory neurons as a function of time. The second row shows the local field potential, which is computed as the mean membrane voltage of inhibitory neurons. The third row shows the evolution of the mean gap junction coupling. The fourth row shows a spectrogram of the network spiking activity, red color represents strong power of the corresponding frequency in the Fourier domain. The last row shows a frequency histogram over the same simulation time.
	(\textbf{A}) With $\nu$ = 40 pA, there is spindle activity. The gap junction coupling oscillating which results in waxing and waning oscillations of around 9 Hz, as shown by the spectrogram and frequency histogram.
	(\textbf{B}) From (A), the TRN external is increased to $\nu$ = 50 pA. The network is in the asynchronous irregular regime. The mean gap junction coupling (third row) is stable and the spectrogram and frequency histogram show no dominating frequency.
	(\textbf{C}) Same as Figure 1E, power in the Fourier domain of the strongest oscillation frequency of the population activity. The dashed line represents the iso-power line of 5 dB.
	(\textbf{D}) Same as Figure 1I, mean spiking activity type. The continuous line represents the iso-activity line of 0.002 and the dashed line represents the iso-power line of 5 dB, extracted from Figure 4C.

    \end{fullwidth}
\end{figure}

\newpage
\begin{figure}[H]
\begin{fullwidth}
	\begin{tabular}{ p{18cm}}
		\resizebox{18cm}{!}{\includegraphics{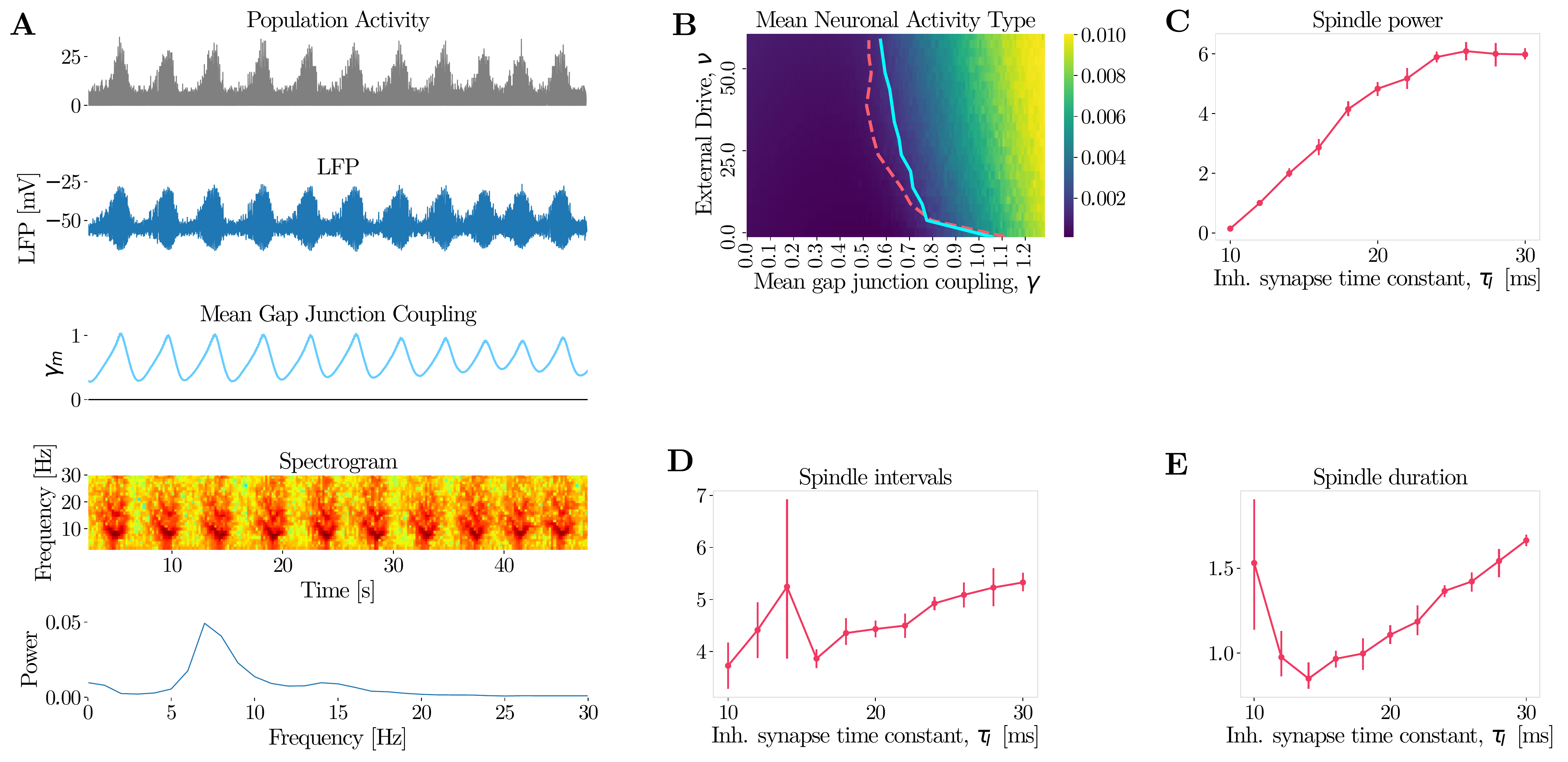}}
	\end{tabular}
	\\
	\caption{ {\bf Potential action of propofol in the generation of spindle activity.} }
	 Panel A consists of the same sub-panels as Figure 3B ($\nu$ = 50 pA) but for a larger value of the time constant of the inhibitory synapses ($\tau_I$ = 20 ms instead of 10 ms). 
	(\textbf{A}) With propofol, the time constants of inhibitory synapses is increased ($\tau_I$ = 20 ms). There is the emergence of spindle activity with waxing and waning oscillations of around 8 Hz, as shown by the spectrogram and frequency histogram.
	(\textbf{B}) Similar activity type diagram as Figure 1E, but for $\tau_I$ = 20 ms. The red dashed line represents an iso-power line of 5 dB, similarly as drawn on panel 4C. The continuous blue line represents an iso-activity line of 0.002. 
	(\textbf{C}) Power of the oscillations during spindle events as function of the inhibitory synapse time constant $\tau_I$. 
	(\textbf{D}) Duration of the spindles oscillations as function of $\tau_I$.
	(\textbf{E}) Intervals between spindle events as function of $\tau_I$.
	
    \end{fullwidth}
\end{figure}

\renewcommand{\thefigure}{S\arabic{figure}}
\renewcommand{\thetable}{S\arabic{table}}
\setcounter{figure}{0}
\setcounter{table}{0}

%
%

\newpage
\section{Supplementary Figures.}

\begin{figure}[H]
\begin{fullwidth}
	\begin{tabular}{ p{16cm}}
		\begin{tabular}{ p{16cm}}
	        		\textbf{\large{A}} \\
				\resizebox{16cm}{!}{\includegraphics{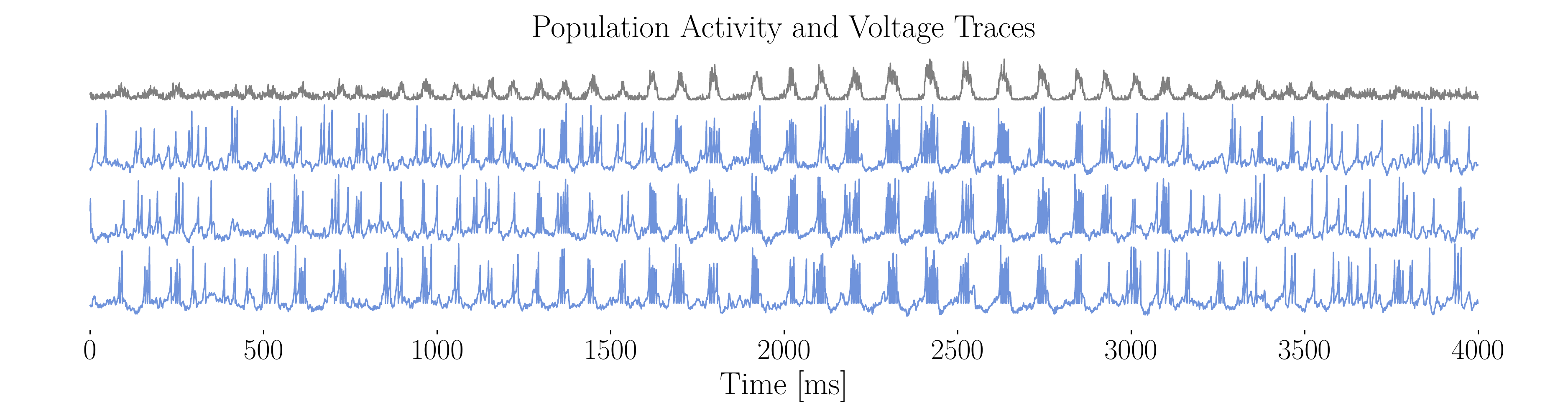}}
				\\
			\textbf{\large{B}} \\
				\resizebox{16cm}{!}{\includegraphics{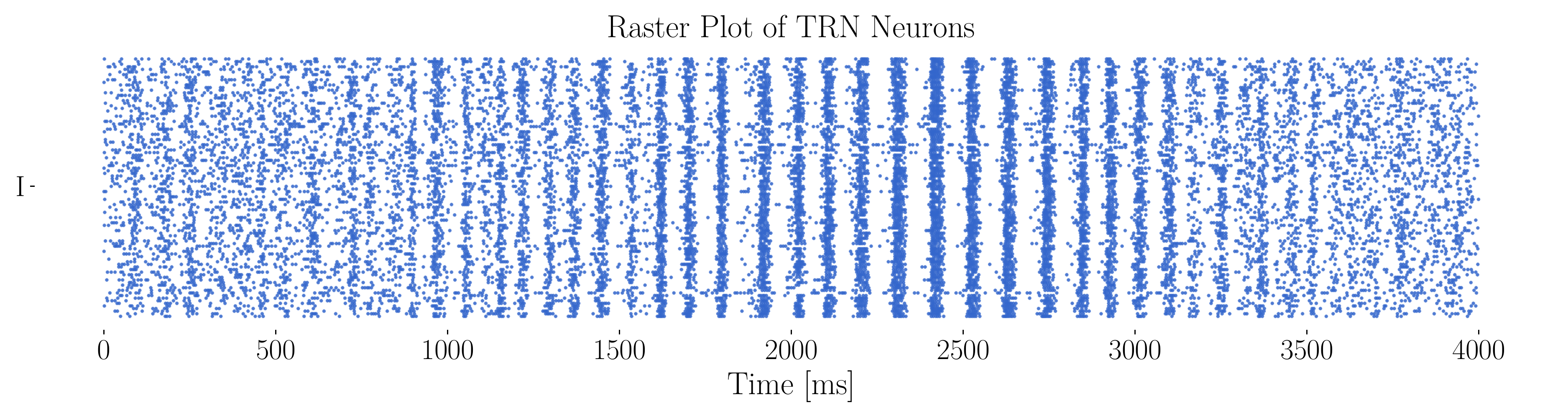}}
				\\
			\textbf{\large{C}} \\
				\resizebox{16cm}{!}{\includegraphics{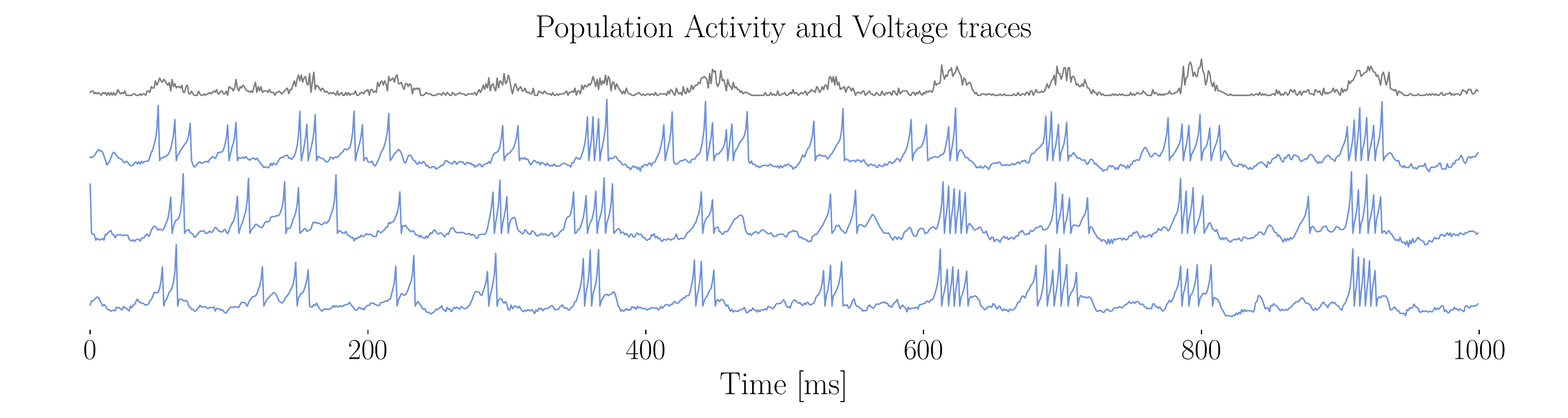}}
		\end{tabular}
	\end{tabular}
	\\
	\caption{ {\bf Voltages traces and network activity during spindle events for the network with just the TRN inhibitory neurons.} }
	(\textbf{A}) In grey, population activity during a spindle event. In blue, voltage traces of 3 different TRN interneurons during the same time. During the spindle, the neuron burst together, but are not synchronized otherwise. 
	(\textbf{B}) Raster plot during the same spindle event as presented in (A). Each line corresponds to one neuron, and the dots represent the spiking times. 
	(\textbf{C}) As for (A), but zoomed-in on only 1000 ms. 
    \end{fullwidth}
\end{figure}

%
%
\begin{figure}[H]
\begin{fullwidth}
	\begin{tabular}{ p{16cm}}
		\begin{tabular}{ p{16cm}}
	        		\textbf{\large{A}} \\
				\resizebox{16cm}{!}{\includegraphics{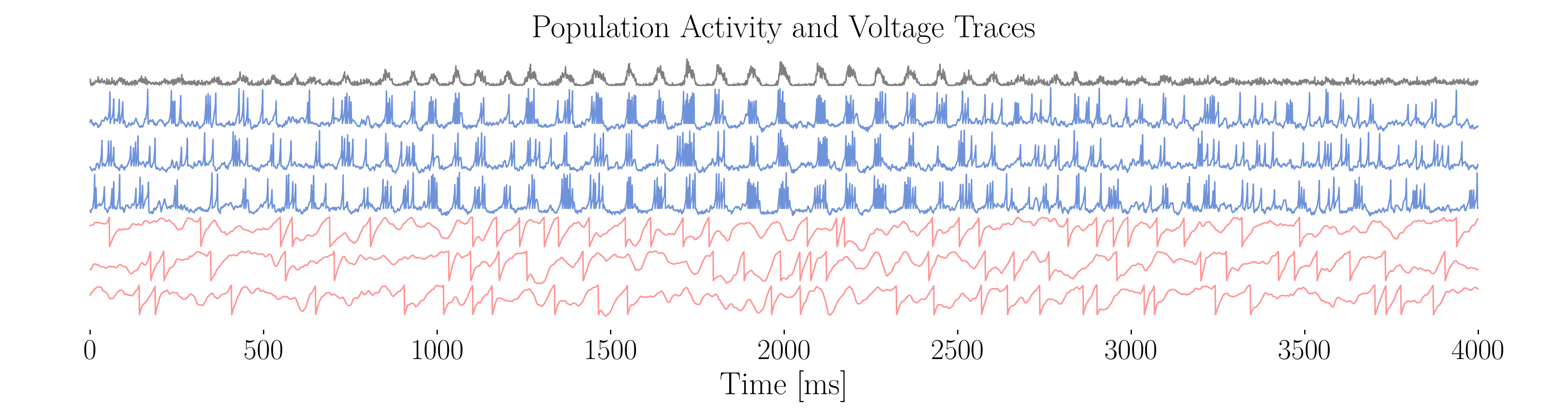}}
				\\
			\textbf{\large{B}} \\
				\resizebox{16cm}{!}{\includegraphics{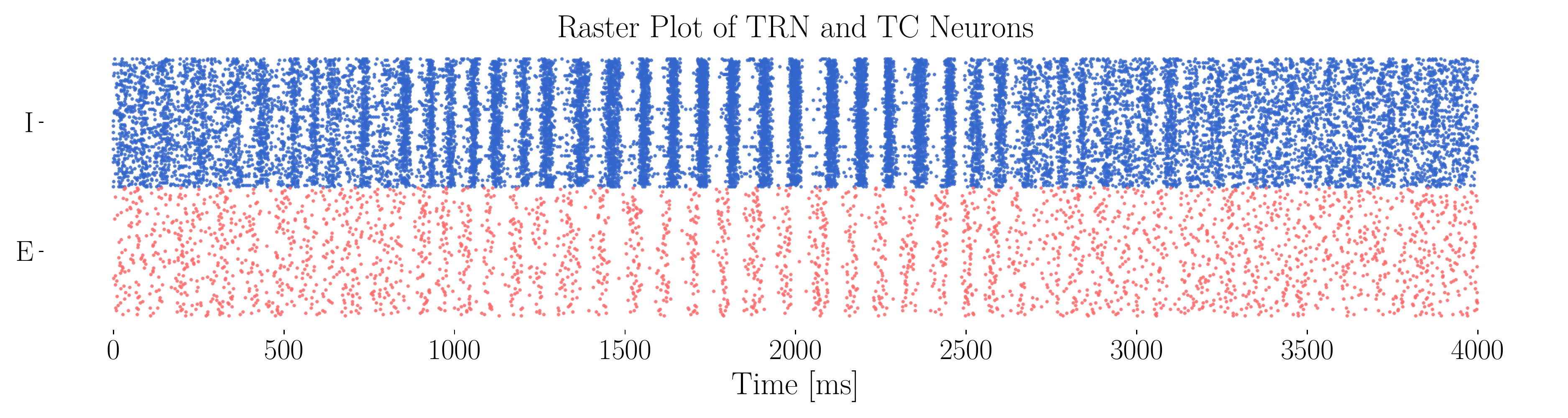}}
				\\
			\textbf{\large{C}} \\
				\resizebox{16cm}{!}{\includegraphics{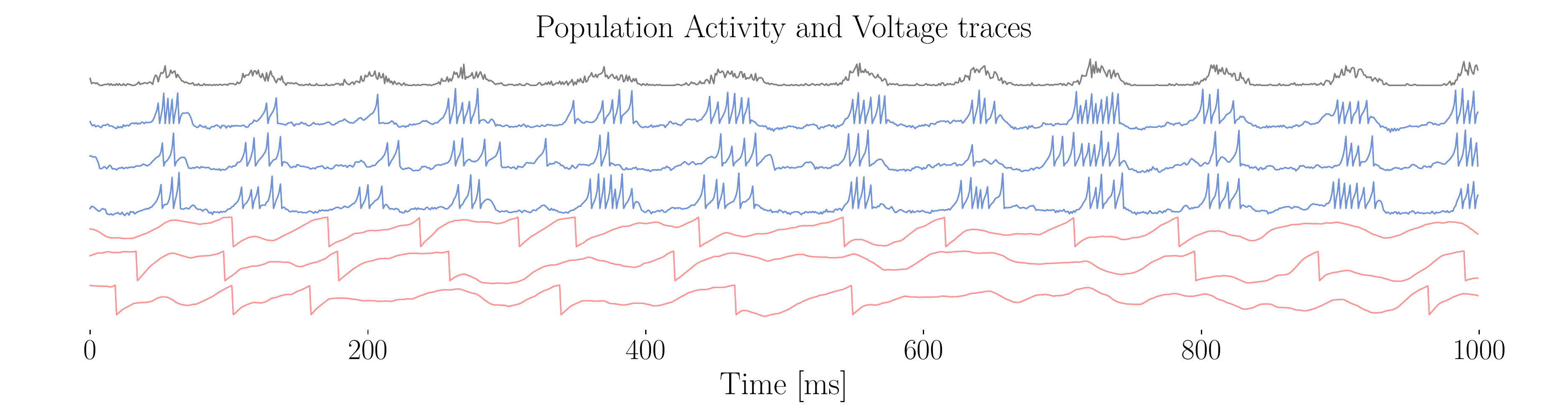}}
		\end{tabular}
	\end{tabular}
	\\
	\caption{ {\bf Voltages traces and network activity during spindle events for the TRN-TC network.} }
	(\textbf{A}) In grey, population activity during a spindle event. Below, voltage traces of different TRN interneurons (blue) and TC excitatory neurons (red) during the same time. During the spindle, the neuron burst together, but are not synchronized otherwise. 
	(\textbf{B}) Raster plot during the same spindle event as presented in (A). The spiking times of 100 TRN interneurons are plotted in blue, and 100 TC neurons in red.
	(\textbf{C}) As for (A), but zoomed-in on only 1000 ms. 
    \end{fullwidth}
\end{figure}

%
%
\newpage
\begin{figure}[H]
\begin{fullwidth}
	\begin{tabular}{ p{16cm}}
		\begin{tabular}{ p{4cm} p{4cm} p{4cm} p{4cm}}

			\textbf{\large{A}} & \textbf{\large{B}} &  \textbf{\large{C}} &  \textbf{\large{D}}   \\
				\resizebox{4cm}{!}{\includegraphics{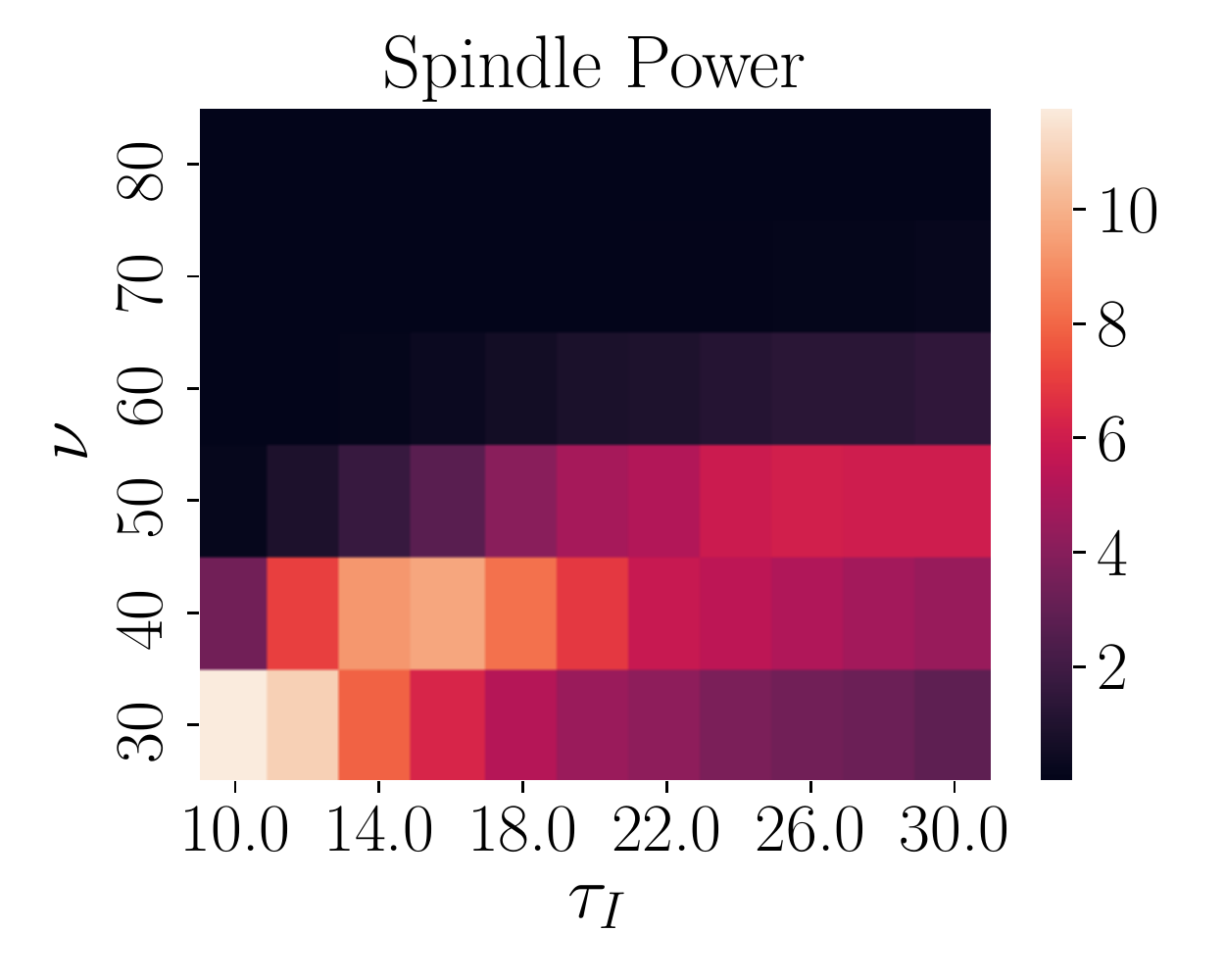}}
	            	&
				\resizebox{4cm}{!}{\includegraphics{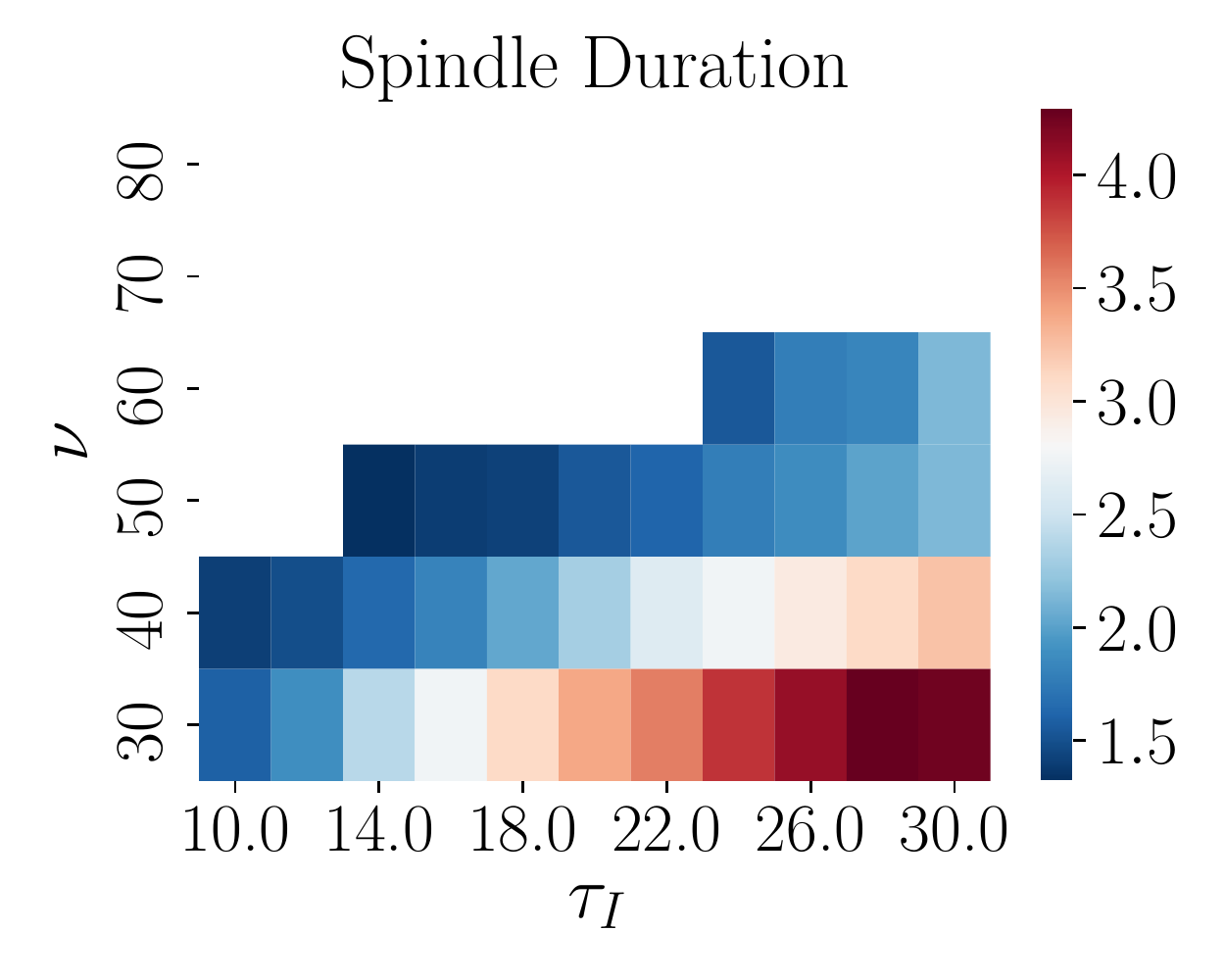}}
				&
				\resizebox{4cm}{!}{\includegraphics{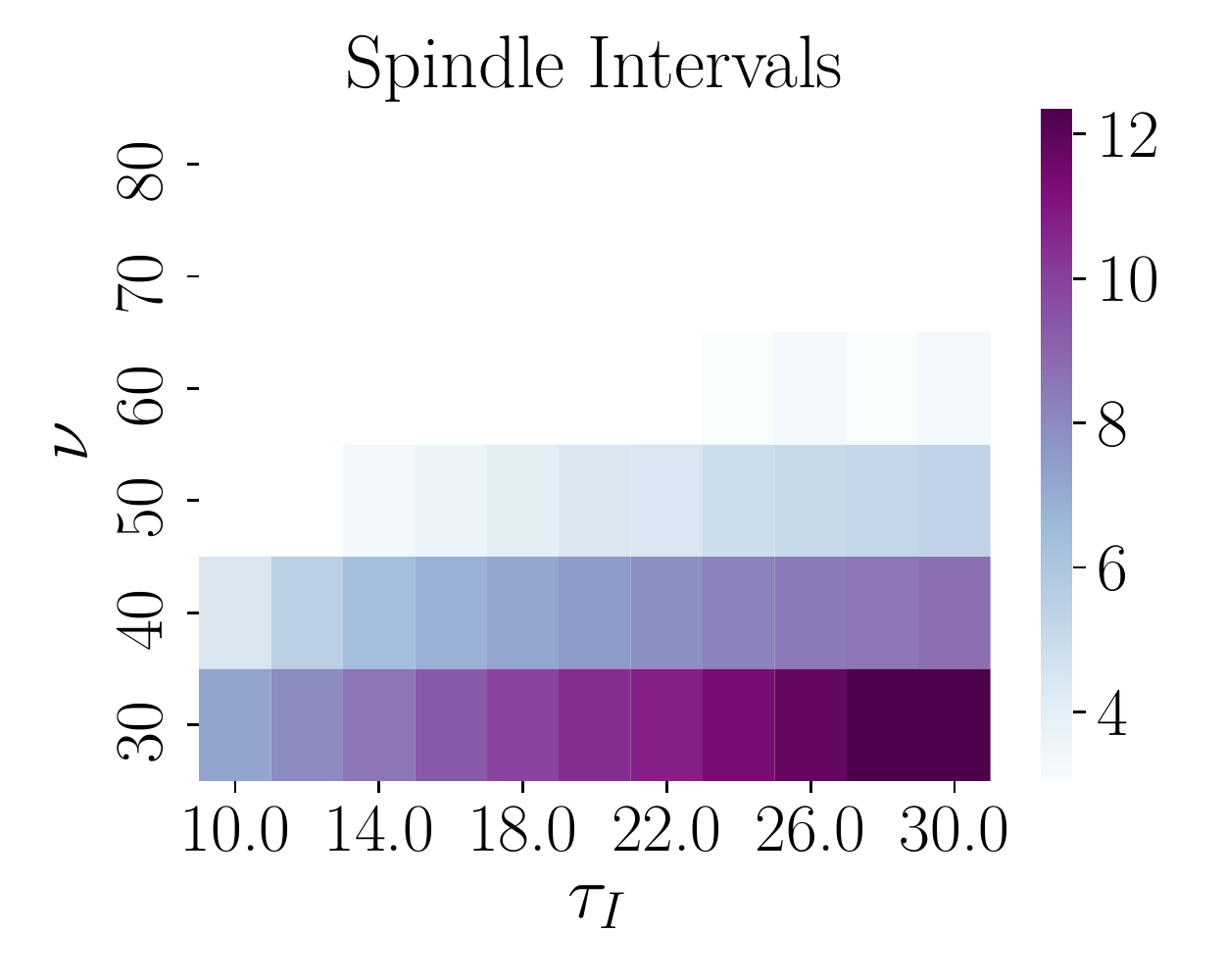}}
				&
				\resizebox{4cm}{!}{\includegraphics{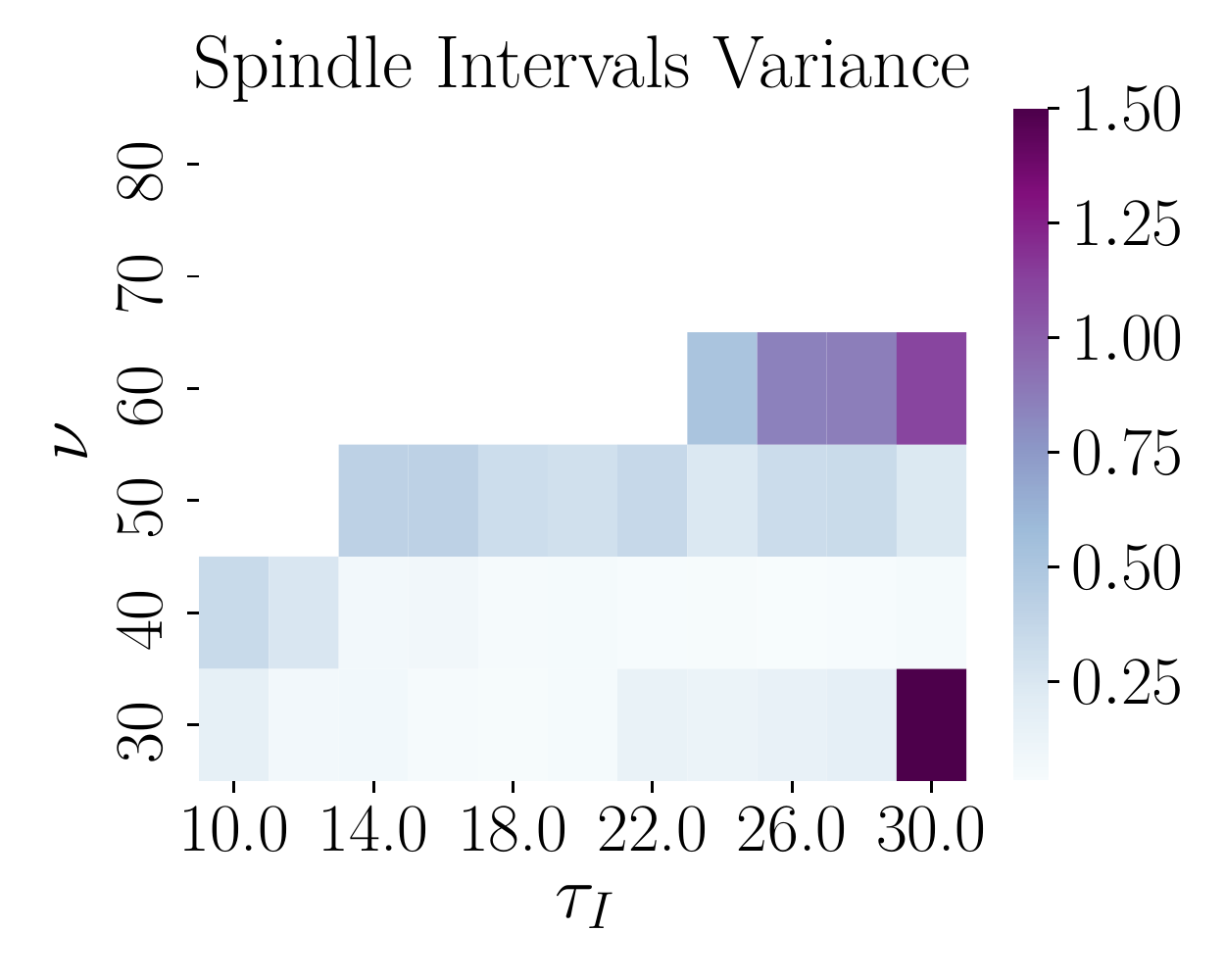}}
				\\

		\end{tabular}
	\end{tabular}
	\\
	\caption{ {\bf Propofol induces spindles through gap junction plasticity.} }
	The x-axis represents the time constant of the TRN interneurons $\tau_I$ and the y-axis represents the external drive to the TRN  $\nu$.
	(\textbf{A}) Mean power of the spindles. Light colors denotes more power. There is spindle activity for high values of $\tau_I$ and $\nu_E$.
	(\textbf{B}) Mean duration of spindle events. 
	(\textbf{C}) Mean time intervals between spindle events.
	(\textbf{D}) Variance in the intervals between spindle events.
         
    \end{fullwidth}
\end{figure}     

\end{document}